\pdfoutput=1
\documentclass[PAPER, atlasdraft=false, cernpreprint, texlive=2025, UKenglish, texmf, orcidlogo]{atlasdoc}
\DeclareOldFontCommand{\sc}{\normalfont\scshape}{\@nomath\sc} %
\usepackage{atlaspackage}
\usepackage{atlasbiblatex}
\usepackage{atlascontribute}
\usepackage{multirow}
\usepackage{atlasphysics}
\usepackage{subfig}
\usepackage{comment}
\usepackage[export]{adjustbox}
\usepackage[tableposition=top]{caption}

\graphicspath{{logos/}{figures/}}

\usepackage{ANA-EXOT-2022-04-PAPER-defs}

\AtlasTitle{Search for neutral long-lived particles that decay into displaced jets in the ATLAS calorimeter in association with leptons or jets using $pp$ collisions at $\sqrt{s}=13$ TeV}

\PreprintIdNumber{CERN-EP-2024-181}
\AtlasAbstract{%

A search for neutral long-lived particles (LLPs) decaying in the ATLAS hadronic calorimeter using 140~fb$^{-1}$ of proton--proton collisions at $\sqrt{s}=13$ TeV delivered by the LHC is presented. The analysis is composed of three channels. The first targets pair-produced LLPs, where at least one LLP is produced with sufficiently low boost that its decay products can be resolved as separate jets. The second and third channels target LLPs respectively produced in association with a $W$ or $Z$ boson that decays leptonically. In each channel, different search regions target different kinematic regimes, to cover a broad range of LLP mass hypotheses and models. No excesses of events relative to the background predictions are observed. Higgs boson branching fractions to pairs of hadronically decaying neutral LLPs larger than 1\% are excluded at 95\% confidence level for proper decay lengths in the range of 30~cm to 4.5~m depending on the LLP mass, a factor of three improvement on previous searches in the hadronic calorimeter. The production of long-lived dark photons in association with a $Z$ boson with cross-sections above 0.1~pb is excluded for dark photon mean proper decay lengths in the range of 20~cm to 50~m, improving previous ATLAS results by an order of magnitude. Finally, long-lived photo-phobic axion-like particle models are probed for the first time by ATLAS, with production cross-sections above 0.1~pb excluded in the 0.1~mm to 10~m range.

}

\author{The ATLAS Collaboration}

\AtlasJournal{JHEP}
\AtlasJournalRef{JHEP 11 (2024) 036}
\AtlasDOI{10.1007/JHEP11(2024)036}

\usepackage{authblk}
\AtlasRefCode{ANA-EXOT-2022-04}

\AtlasNote{ANA-EXOT-2022-04-PAPER}

\hypersetup{pdftitle={ATLAS document},pdfauthor={The ATLAS Collaboration}}

\begin{document}

\maketitle

\tableofcontents

%
%
%

%

\section{Introduction}
\label{sec:intro}

Many extensions to the Standard Model (SM) of particle physics predict new particles whose decays are suppressed by weak coupling constants, small mass differences between particles, or heavy mediators.
Such particles can acquire large lifetimes, becoming long-lived particles (LLPs).
LLPs are featured in many well-motivated theories including various supersymmetry models~\cite{ArkaniHamed:2012gw,Giudice:1998bp,Barbier:2004ez,Csaki:2015fea,Fan:2011yu,Fan:2012jf};
neutral naturalness models~\cite{Chacko:2015fbc,Burdman:2006tz,Cai:2008au,Chacko:2005pe} featuring a Hidden Sector (HS)~\cite{Strassler:2006im,Strassler:2006ri,Chan:2011aa} that addresses the hierarchy problem;
models that seek to incorporate dark matter~\cite{Baumgart:2009tn,Kaplan:2009ag,Dienes:2011ja,Dienes:2012yz}, or explain the matter–antimatter asymmetry of the universe~\cite{Cui:2014twa};
and models including heavy neutrinos~\cite{Helo:2013esa,Batell:2016zod} that provide an explanation for the origin of light neutrino masses and mixings.
Axion-like particles (ALPs), which may help resolve the strong CP problem, are also predicted to be long-lived in some parts of the parameter space~\cite{Brivio:2017ije}.
Not only could LLPs provide a novel route to a groundbreaking discovery, but they can also probe regions of common benchmark models~\cite{Haisch_2023}, such as the two-Higgs doublet model that is widely used to explore dark matter topologies, in regions inaccessible to the wider ATLAS search programme involving prompt particles.

This paper studies electrically neutral LLPs decaying hadronically in the hadronic calorimeter of the ATLAS detector at the Large Hadron Collider (LHC). This corresponds to particles with a lifetime ($\tau$) times the speed of light ($c$) between a few centimeters and tens of meters. Since the LLPs are neutral,  they do not leave any hits in the ATLAS tracking system. Their decay into quarks or gluons, which later hadronise, leads to jets in the calorimeters. For LLPs that decay after the electromagnetic part of the calorimeter, the distribution of energy of the resulting jets differs from that of standard jets, as they have a very low electromagnetic component. This fact can be used to identify displaced jets from SM jets and other sources of background. The ``CalRatio'' quantity, representing the ratio of the energy in the hadronic calorimeter to the energy in the electromagnetic calorimeter, can also serve this purpose.

There exist several previous searches for pair-produced neutral LLPs decaying hadronically at the LHC\@. The search for displaced jets at LHCb in Ref.~\cite{LHCb:2017xxn} is sensitive to $c\tau$ values from $\sim$1~mm to $\sim$0.1~m. The most recent searches by the CMS Collaboration at $\sqrt{s}=13$ \TeV~\cite{CMS-EXO-19-021,CMS-EXO-20-003,CMS-EXO-22-020-new} involve jets with displaced vertices in the tracking system or muon detectors~\cite{CMS-EXO-21-008}, and are sensitive to $c\tau$ values from $\sim$1 mm to $\sim$10~m. Previous ATLAS searches at $\sqrt{s}=13 \TeV$ looked for displaced vertices in the tracking system~\cite{HIGG-2017-05,EXOT-2018-57, HDMI}, hadronic calorimeter~\cite{EXOT-2019-23}, pairs of reconstructed vertices in the muon spectrometer~\cite{EXOT-2017-05}, or the combination of one displaced vertex in the muon spectrometer and one in the inner tracking detector~\cite{EXOT-2018-61}. A search for hadronic decays of LLPs in association with a $Z$ boson was performed by ATLAS in Ref.~\cite{EXOT-2017-24} using early Run 2 data. These ATLAS searches are complementary, and together provide coverage of $c\tau$ values extending from effectively prompt to $\sim$200~m.

Previous searches for displaced jets in the calorimeter at ATLAS have found limitations at the level of the trigger. Indeed, if the LLPs are produced without any additional objects (as happens for example for gluon--gluon fusion production of a Higgs boson-like mediator decaying into LLPs), the decay products can often fail to meet minimum energy thresholds required to record the event.  This is particularly true in low-mass and low-boost kinematic regimes. The analysis presented in this paper differs from its predecessors insofar as it searches for LLPs decaying into displaced jets in addition to other objects in the event, such as the decay products of $W$ or $Z$ bosons, or resolved jets from another LLP\@. These differences allow sensitivity to lower decay lengths or to trigger on prompt objects, gaining in trigger efficiency at the cost of a lower production cross-section. The analysis then uses the same displaced jet identification technique as in Ref.~\cite{EXOT-2019-23} to select candidate jets resulting from LLP decays in the calorimeter.
These will have large values of the CalRatio variable defined above.

The analysis is split into three channels. The CalRatio + two jets (CalR+2J) channel targets LLPs resulting from gluon--gluon fusion production of the mediator particle in an HS model, leading to two LLPs in the event. One LLP decays in the calorimeter, with its decay products merged into a single displaced jet. The other LLP decays with a short enough decay length and low enough boost to produce two resolved jets from its decay products.

The CalRatio + $W$ boson (CalR+$W$) and CalRatio + $Z$ boson (CalR+$Z$) channels target displaced jets produced in association with a SM vector boson. This results, for example, from an HS mediator produced with a SM vector boson or an ALP radiated from a vector boson. In each case, the vector boson's leptonic decay is exploited.

These different final states require different event selections and triggering strategies. All three channels use the same displaced jet tagger~\cite{EXOT-2019-23}. The main backgrounds for the CalR+2J channel are SM multijets, and non-collision backgrounds (NCBs) such as beam-induced background (BIB) and cosmic rays (although these are ultimately reduced to a negligible level in all channels). The CalR+$W$ and CalR+$Z$ channels have main backgrounds originating from SM processes involving vector bosons produced with jets, and single- or pair-production of top quarks. In all channels, the final background estimate is data-driven (using the likelihood-based ABCD method).

The rest of this document is organised as follows. The ATLAS detector is described in Section~\ref{sec:detector}. The datasets and simulated samples used for the analysis are reported in Section~\ref{sec:datasets}. The definitions of the reconstructed objects are provided in Section~\ref{sec:obj_selection}.
The selection criteria applied to define each channel are laid out in Section~\ref{sec:event_selection}. The description of the relevant background sources and how they are evaluated is available in Section~\ref{sec:bkgs}. Systematic uncertainties are listed in Section~\ref{sec:systematics}. Finally, the statistical analysis and results are reported in Section~\ref{sec:results} before the conclusions in Section~\ref{sec:conclusions}.


\section{The ATLAS detector}
\label{sec:detector}

The ATLAS detector~\cite{PERF-2007-01} at the LHC covers nearly the entire solid angle around the collision point.\footnote{ATLAS uses a right-handed coordinate system with its origin at the nominal interaction point (IP)
in the centre of the detector and the \(z\)-axis along the beam pipe.
The \(x\)-axis points from the IP to the centre of the LHC ring,
and the \(y\)-axis points upwards.
Polar coordinates \((r,\phi)\) are used in the transverse plane,
\(\phi\) being the azimuthal angle around the \(z\)-axis.
The pseudorapidity is defined in terms of the polar angle \(\theta\) as \(\eta = -\ln \tan(\theta/2)\) and is equal to the rapidity
$ y = \frac{1}{2} \ln \left( \frac{E + p_z c}{E - p_z c} \right) $ in the relativistic limit.
Angular distance is measured in units of \(\Delta R \equiv \sqrt{(\Delta y)^{2} + (\Delta\phi)^{2}}\).}

It is a multipurpose detector consisting of an inner tracking detector surrounded by a thin superconducting solenoid, electromagnetic and hadronic calorimeters, and a muon spectrometer incorporating three large superconducting toroidal magnets. The inner detector system is immersed in a $2$~T axial magnetic field and provides charged-particle tracking in the range of $|\eta| < 2.5$.

The high-granularity silicon pixel detector covers the vertex region and typically provides four measurements per track. The layer closest to the interaction point is known as the insertable B-Layer~\cite{ATLAS-TDR-2010-19,PIX-2018-001}. The pixel detector is surrounded by the silicon microstrip tracker, which usually provides four three-dimensional measurement points per track. These silicon detectors are complemented by the transition radiation tracker, with coverage up to $|\eta| = 2.0$.

The calorimeter system covers the pseudorapidity range $|\eta| < 4.9$. Within the region $|\eta| < 3.2$, electromagnetic calorimetry is provided by barrel and endcap high-granularity lead/liquid-argon (LAr) electromagnetic calorimeters (together referred to as the ECAL), with an additional thin LAr presampler covering $|\eta| < 1.8$ to correct for energy loss in material upstream of the calorimeters. The ECAL extends from 1.5~m to 2.0~m in radial distance $r$  in the barrel and from 3.6~m to 4.25~m in $|z|$ in the endcaps. Hadronic calorimetry is provided by a steel/scintillator-tile calorimeter (HCAL), segmented into three barrel structures within the range $|\eta| < 1.7$, and two copper/LAr hadronic endcap calorimeters covering $|\eta| > 1.5$. The HCAL covers the region from 2.25~m to 4.25~m in $r$ in the barrel (although the HCAL active material extends only up to 3.9~m) and from 4.3~m to 6.05~m in $|z|$ in the endcaps. The solid angle coverage is completed with forward copper/LAr and tungsten/LAr calorimeter modules optimised for electromagnetic and hadronic measurements, respectively.

The calorimeters have a highly granular lateral and longitudinal segmentation. Including the presamplers, there are seven sampling layers in the combined central calorimeters (the LAr presampler, three in the ECAL barrel and three in the HCAL barrel), and the endcap regions provide up to eight sampling layers (the presampler, three in ECAL endcaps and four in HCAL endcaps). The forward calorimeter modules provide three sampling layers in the forward region. The combined depth of the calorimeters for hadronic energy measurements is more than nine hadronic interaction lengths nearly everywhere across the full detector acceptance.

The muon spectrometer comprises separate trigger and high-precision tracking chambers measuring the deflection of muons in the magnetic field generated by the superconducting air-core toroids. The field integral of the toroids ranges between 2.0 and 6.0~Tm across most of the detector.

The ATLAS detector records events using a tiered trigger system~\cite{TRIG-2016-01}. The level-1 (L1) trigger is implemented in custom electronics and reduces the event rate from the LHC crossing frequency of 40~MHz to a design value of 100~kHz. The second level, known as the high-level trigger (HLT), is implemented in software running on a commodity PC farm that processes the events and reduces the rate of recorded events to 1~kHz.
A software suite~\cite{SOFT-2022-02} is used in data simulation, in the reconstruction and analysis of real and simulated data, in detector operations, and in the trigger and data acquisition systems of the experiment.


\section{Data and simulated event samples}
\label{sec:datasets}

The analysed dataset was collected between 2015 and 2018 by the ATLAS detector from proton--proton ($pp$) collisions at $\sqrt{s}=13$ \TeV at the LHC\@.
The total integrated luminosity of this dataset is 140~fb$^{-1}$.
Data where the LHC beams were not stable or not all subdetectors were operational were excluded~\cite{DAPR-2018-01}.

For the CalR+2J channel, dedicated LLP triggers (dubbed ``CalRatio triggers'' as they exploit the ratio of energy in the electromagnetic and hadronic parts of the calorimeter) were employed. They were detailed Ref.~\cite{EXOT-2019-23}, and a short summary is given here. At L1, jets were reconstructed in $0.2\times 0.2$ regions in the $\eta ~ \times ~ \phi$ plane, motivated by the fact that signal jets tend to be narrower than regular SM jets. Events with jets with transverse energy \et above 60 or 100 \GeV (depending on the data-taking period) are selected.
In some run periods, triggers were available with the requirement that L1 jet objects were accepted only if there existed no deposit in the electromagnetic calorimeter in the same $\eta-\phi$ position as the jet. This requirement acted as a proxy to the CalRatio quantity, allowing the \et thresholds to be reduced to 30 \GeV. At HLT, the triggering jet was additionally required to satisfy $|\eta| < 2.5$ where tracking information is available. Further, the jet should have a high proportion of its energy in the HCAL, satisfy a modified noise-suppression selection and satisfy a BIB-removal algorithm exploiting, for example, the timing and alignment of deposits in $\phi$. In the following, the dataset collected with the CalRatio triggers is referred to as the ``main'' dataset.

In the remaining channels, single or dilepton triggers~\cite{TRIG-2018-05,TRIG-2018-01} were employed, where only triggers running unprescaled for the whole data-taking period in a given year were considered.

Three additional datasets were collected for the study of NCBs and construction of validation and control regions. These datasets are the same as described in Ref.~\cite{EXOT-2019-23} and are summarised briefly here. The BIB dataset was collected from events satisfying all the requirements of the CalRatio triggers but failing to pass the BIB-removal algorithm. The cosmic ray dataset was collected using the same trigger selection but from events recorded during empty bunch crossings. Finally, a dijet dataset was selected using a single-jet-based trigger~\cite{TRIG-2012-01} and vetoing on the CalRatio triggers to make it orthogonal to the main dataset. This dataset is used to test the effect of the modelling in the displaced jet neural network tagger's output.

The main SM backgrounds are SM multijets in the CalR+2J channel, and W/Z+jets, $t\bar{t}$ and single-top-quark events in the CalR+$W$ and CalR+$Z$ channels. Although data-driven methods are used to do the final background estimations, Monte Carlo (MC) simulated events are needed to train the machine-learning (ML) discriminants and evaluate certain systematic uncertainties. The SM multijet samples were generated at leading order (LO) with \textsc{Pythia 8.186}~\cite{Sjostrand:2007gs} using the A14 set of tuned parameters (tune)~\cite{ATL-PHYS-PUB-2014-021} for parton showering and hadronisation. The \textsc{NNPDF2.3LO} parton distribution function (PDF) set~\cite{Ball:2012cx} was used. Events containing a $W$ or $Z$ boson with jets were simulated with \textsc{Sherpa v2.2.1}~\cite{Gleisberg:2008ta,Bothmann:2019yzt}. Next-to-leading-order (NLO) accurate matrix elements for up to two jets, and LO-accurate matrix elements for up to four jets were calculated with the \textsc{Comix}~\cite{Gleisberg:2008fv} and \textsc{OpenLoops}~\cite{Cascioli:2011va,Denner:2016kdg} libraries. The default \textsc{Sherpa} parton shower~\cite{Schumann:2007mg} was used. The production of $t\bar{t}$ events was modelled using the \textsc{Powheg Box v2}~\cite{Frixione:2007nw,Nason:2004rx,Frixione:2007vw,Alioli:2010xd} generator at NLO with the NNPDF3.0NLO PDF set and the \texttt{hdamp} parameter set to 1.5 times the top quark mass~\cite{ATL-PHYS-PUB-2016-020}. The events were interfaced with \textsc{Pythia 8.230}~\cite{Sjostrand:2014zea} using the A14 tune and the \textsc{NNPDF2.3LO} PDF set. The NLO $t\bar{t}$  inclusive production cross-section was corrected to the theory prediction at next-to-next-to-leading-order (NNLO) in QCD including the resummation of next-to-next-to-leading logarithmic (NNLL) soft-gluon terms calculated using Top++2.0~\cite{Beneke:2011mq,Cacciari:2011hy,Baernreuther:2012ws,Czakon:2012zr,Czakon:2012pz,Czakon:2013goa,Czakon:2011xx}.
Single-top-quark production was modelled using \textsc{Powheg Box v2}~\cite{Re:2010bp,Nason:2004rx,Frixione:2007vw,Alioli:2010xd} at NLO in QCD in the five-flavour scheme with the NNPDF3.0NLO PDF set~\cite{Ball:2014uwa}. The diagram-removal scheme~\cite{Frixione:2008yi} was employed to address the interference with $t\bar{t}$  production~\cite{ATL-PHYS-PUB-2016-020}. The events were interfaced with \textsc{Pythia 8.230} using the A14 tune and the \textsc{NNPDF2.3LO} PDF set. The inclusive cross-section was corrected to the theory prediction calculated at NLO in QCD with NNLL soft-gluon corrections~\cite{Aliev:2010zk,Kant:2014oha}.

Three types of benchmark signals are considered. The first is the HS model~\cite{Chung2006,Chang2008,Strassler:2006im,Strassler:2006ri}, where a scalar boson $\Phi$ (the Higgs boson, or a lighter or heavier particle that behaves similarly) acts as mediator between the SM and the HS\@. The $\Phi$ can decay into neutral long-lived scalars, denoted $S$, which are the LLPs. The scalars decay chiefly into the heaviest kinematically accessible fermion pairs: typically $b$-quarks in most of the considered signals. This is the same model as studied in Ref.~\cite{EXOT-2019-23}. Events were simulated using \textsc{MadGraph5\_aMC@NLO v2.6.2}~\cite{MadGraphNLO} at LO with the NNPDF2.3LO PDF set. Two production modes for $\Phi$ are considered: gluon--gluon fusion production for the CalR+2J channel and associated vector boson ($W$ or $Z$) production for the CalR+$W$/$Z$ channels. Vector boson fusion (VBF) is not considered: in the CalR+2J channel all objects that play a role in the selection come from LLP decays. Hence, the efficiencies for events produced from gluon--gluon fusion and VBF productions are expected to be similar. Given the relative cross-sections between these production modes, no sensitive additional contribution is expected from the VBF production. The HS samples produced with a $W$ or $Z$ boson are referred to as $W$HS and $Z$HS, respectively. For the gluon--gluon fusion samples, the $\Phi$ transverse momentum distribution was reweighted to match NLO predictions using the same event generator. Cross-sections of 48.6~pb for gluon--gluon fusion, 0.45~pb for associated production with a $W$ boson decaying leptonically, and 0.09~pb for associated production with a $Z$ boson decaying leptonically (all extracted from the NNLO calculation~\cite{Cepeda:2019klc}) are assumed when normalising results for the case where the mediator is the SM Higgs boson. Several sets of samples were generated, with different assumptions for particle masses, in the range of $60$ to $1000$ \GeV for the mediator, and $5$ to $475$ \GeV for the long-lived scalar.
The second type of signal model, considered in the CalR+$W$/$Z$ channels, contains photo-phobic ALPs~\cite{Brivio:2017ije} that are radiated from vector bosons and decay into gluons. The ALP masses vary between $0.1$ and $40$ \GeV, and the coupling of the ALP to gluons, which determines its lifetime, varies between $10^{-7}$ and $10^{-2}$. In this model, only vector boson decays into electrons or muons are considered. The ALP samples were generated with \textsc{MadGraph5\_aMC@NLO v2.9.3} at LO with the NNPDF2.3LO PDF set. The samples are referred to as $Z$ALP and $W$ALP\@.
The third benchmark sample is a long-lived dark photon ($Z_d$) model~\cite{PhysRevD.85.115019,PhysRevD.88.015022}, where the $Z_d$ is produced with a $Z$ during the decay of a scalar mediator. It is referred to here as the $HZZ_d$ model. This model was studied by ATLAS in Ref.~\cite{PhysRevLett.122.151801}. Mediator masses between 250 and 600 \GeV are considered, with dark photon masses ranging between 5 and 400 \GeV. The samples were generated with \textsc{MadGraph5\_aMC@NLO2.6.7} at LO with the NNPDF2.3LO PDF set. For all signal samples, parton showering and hadronisation were modelled using \textsc{Pythia8} with the A14 tune. This model was last searched for by ATLAS in the $Z$ + displaced jet channel in Ref.~\cite{EXOT-2017-24}.

\begin{figure}[ht!]
\centering
\subfloat[]{
\includegraphics[width=0.3\textwidth]{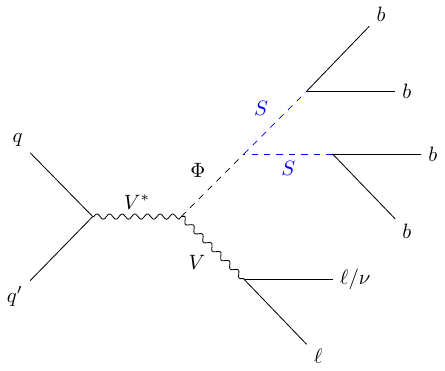}
\label{fig:feynmanA}
}
\hspace{\stretch{1}}
\subfloat[]{
\includegraphics[width=0.3\textwidth]{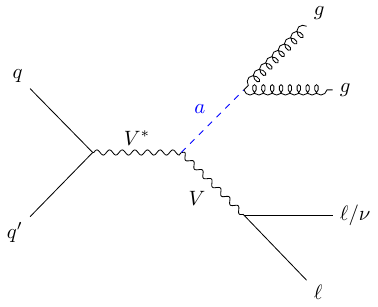}
\label{fig:feynmanB}
}\hspace{\stretch{1}}
\subfloat[]{
\includegraphics[width=0.3\textwidth]{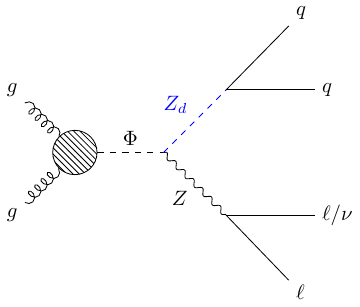}
\label{fig:feynmanC}
}

\caption{Example Feynman diagrams for the signal processes explored in this paper. (a) Model where the mediator ($\Phi$) is produced in association with a vector boson $V=W$ or $Z$, and the LLPs $S$ decay into fermions (mainly $b$-quarks). (b) ALP model where the long-lived ALP $a$ decays into gluons, and is produced in association with a vector boson. (c) Dark photon $Z_d$ model produced with a $Z$ boson from the decay of a mediator $\Phi$.}
\label{fig:feynman}
\end{figure}

Example Feynman diagrams for these three types of model are shown in Figure~\ref{fig:feynman}.
For all signal samples, the generated LLP mean proper lifetime ($\tau_\textrm{gen}$) was chosen to maximise the fraction of decays in the ATLAS hadronic calorimeter and muon system (these mean proper lifetimes are typically of the order of a few meters, but vary across samples due to time dilation effects). The effect of multiple interactions per bunch crossing (pile-up) was modelled by overlaying the simulated hard-scattering event with inelastic $pp$ collision events generated with \textsc{Pythia 8.186} using the \textsc{NNPDF2.3LO} PDF set and the A3 tune~\cite{ATL-PHYS-PUB-2016-017}.
The detector geometry and response were simulated with \textsc{Geant4}~\cite{SOFT-2010-01, Geant4}. The standard ATLAS reconstruction software is used for both the simulation and collision data.


\section{Event reconstruction}
\label{sec:obj_selection}

Three types of reconstructed objects are used: jets, leptons and missing transverse momentum. Hadronically decaying LLPs will form displaced jets, which are the common signature in all the channels. The leptons and missing transverse momentum are exploited to reconstruct $W$ and $Z$ boson candidates in the CalR+$W$ and CalR+$Z$ channels. The formal definitions of each object type are described below.

\textbf{Jets:} Jets are reconstructed using the anti-k$_t$ algorithm~\cite{Cacciari:2008gp, Fastjet} with a radius parameter $R=0.4$ using calorimeter energy clusters only (tracking information is not used). Jets are further calibrated to account for the predicted detector response in MC simulation, and a residual calibration of jets in data is derived through \emph{in situ} measurements~\cite{JETM-2018-05}. ``Clean jets'' are defined as those that have transverse momentum ($\pt$) above 40 GeV, $|\eta| < 2.5$ and satisfy the CalRatio jet cleaning requirement described in Ref.~\cite{EXOT-2019-23}.  A ``trackless jet'' is defined as a clean jet with $\Delta R_\textrm{min}  > 0.2$, where $\Delta R_\textrm{min} $ represents the angular distance between the jet axis and the closest track with $\pt >2$ \GeV.
SM jets, which typically have multiple tracks in their cone, will have small values of this quantity, while anomalous trackless jets will yield large values.
The $\sum\Delta R_\textrm{min}$ is defined as the sum of $\Delta R_\textrm{min} $ over jets above a certain \pt threshold, which is optimised differently according to the channel: 40 \GeV for the CalR+$W$/$Z$ channels and 50 \GeV for the CalR+2J channel. This is a useful discriminant to select events containing displaced jets.
This quantity was found to be stable across pile-up values encountered during LHC Run 2.
The term ``displaced jet'' is generically used to refer to jets that originated away from the interaction point: in the context of this analysis, it is taken to mean jets that began in the ATLAS calorimeters.
A CalRatio jet candidate must be matched (within a cone of radius 0.2) to an HLT jet that met the criteria of one of the CalRatio triggers. It must also satisfy $\Delta R_\textrm{min} (\textrm{jet, track}) > 0.2$ and $\log_{10} (E_\textrm{H} / E_\textrm{EM}) > 1.2$, where \lograt is the base-10 logarithm of their CalRatio value.

All analysis channels make use of a neural network (NN) classifier trained to distinguish signal-like jets (arising from a LLP decay) from BIB-like jets and SM multijets. This displaced NN jet tagger was originally trained for the analysis in Ref.~\cite{EXOT-2019-23}, which contains all the details about its architecture, training and performance. Only a brief summary of the key points is included here. The architecture chosen was a set of convolutional layers fed into a long short-term memory (LSTM) layer~\cite{LSTM}. An additional adversarial component was added to prevent the network from exploiting mismodelling of some of the features (further details are provided in Section~\ref{sec:systematics}). The inputs to the NN are low-level features of the jets from the tracker (positions, momenta, impact parameter and quality variables of inner-detector tracks that are within $\Delta R = 0.2$ of jets), calorimeters (the fraction of the jet energy deposited in each layer of the ECal and HCal and momenta, timing information and positions of calorimeter topoclusters, i.e. collections of calorimeter cells used in jet reconstruction~\cite{PERF-2014-07} associated with each jet) and muon system (spatial and timing information for muon track segments within $\Delta \phi = 0.2$ of a jet). The NN was trained on SM multijet MC samples and the BIB dataset, and a combination of HS signal samples. Two trainings were provided: one using signals with mediator masses below or equal to 200 \GeV; the other using signal with mediator masses strictly above 200 \GeV. These two versions of the NN are referred to as the \emph{low-\ET} and \emph{high-\ET} NNs respectively. Each NN outputs three scores for each jet, which relate to the probability that such a jet would be the result of an LLP decay, a BIB hit, or a non-displaced jet. These scores are referred to as signal-score, BIB-score and multijet-score, and sum to unity. The modelling of the input features is checked using the dijet dataset described in Section~\ref{sec:datasets}. A dedicated procedure is used to evaluate the residual uncertainty in the performance of the NN due to potential mis-modelling of the input variables, as described in Section~\ref{sec:systematics}.
The per-jet NN is then exploited in the three analysis channels, which are described separately in Section~\ref{sec:event_selection}, selecting the jets with the highest signal-scores.

\textbf{Leptons:}  ``Baseline'' leptons ($e$ or $\mu$) are required to have \pt above 10 \GeV, and satisfy $|\eta|<2.47$ for electrons and $|\eta|<2.5$ for muons. Electrons in the gap between the barrel and the endcap are excluded. Further, these leptons are expected to meet the standard ATLAS requirements on track-to-vertex association, \textit{Medium} identification and \textit{Tight} isolation conditions, as defined in Refs.~\cite{EGAM-2018-01,MUON-2018-03}. The standard lepton reconstruction efficiency correction factors are used.
A procedure is implemented to resolve  ambiguities arising from the fact that electrons, muons and jets are identified and reconstructed independently.  Only leptons satisfying the baseline selections are considered for this process, and calibrated jets used for the displaced jet identification. First, if two electrons share a track in the inner detector, then the lepton with the lowest \pt is discarded (in practice this is a negligible effect). Next if a muon and electron share a track, then the electron is discarded unless the muon was seeded from calorimeter energy clusters. Any jets within $\Delta R = 0.3 $ of   electrons or muons are removed from the final state. Then, any selected electrons or muons within $0.2 < \Delta R < 0.4$ of any remaining selected jet are also removed.

\textbf{Missing transverse momentum and vector bosons:}
The missing transverse momentum, $\Vec{p}_\textrm{T}^\textrm{miss}$, is defined as the negative vector sum of the transverse momenta of all baseline electrons, muons and jets in the event, plus an additional soft term corresponding to tracks not associated with any identified lepton or jet~\cite{JETM-2020-03}. Its magnitude, \met, is referred to as the missing transverse energy.
$W$ boson candidates are formed in events containing exactly one baseline lepton (denoted by $\ell$) matched to a lepton trigger object with $\pt > 27 \gev$ and $\met > 30\GeV$.
The transverse invariant mass of the system, given by $\sqrt{2\met \pt^\ell (1-\cos(\Delta\phi(\met, \ell)))}$, is required to be above 50 \GeV to remove multijet background contamination.
$Z$ boson candidates are defined in events containing at least one same-flavour opposite-sign pair of baseline leptons, each with the leading and subleading leptons satisfying $\pt> 25 \GeV$ and $\pt> 10 \GeV$ respectively.
At least one of the selected leptons is required to match to a lepton trigger object. The invariant mass of the dilepton system $m_{\ell\ell}$ is required to be in the range of  60 $ < m_{\ell\ell} <$ 120~\GeV. If multiple lepton pairs are possible, only the pair  with the invariant mass closest to the $Z$ boson mass is kept. The lepton $\pt$ thresholds used in the construction of vector boson candidates were selected following an optimisation process with sensitivity to LLP signals as the figure of merit.


\section{Event selection}
\label{sec:event_selection}

Events are required to first satisfy at least one of the triggers listed in Section~\ref{sec:datasets}.
In the CalR+$W/Z$ channels, the logical ``OR'' of the unprescaled triggers with the lowest \pt thresholds is utilised.
Adding the CalRatio triggers into the ``OR' did not result in a significant gain in signal efficiency, so they were not included.
In both the channels, the trigger efficiency increases with lepton \pt, and it reaches a plateau for values of \pt above 50 \GeV (in the CalR+$Z$ channel) and 100 \GeV (in the CalR+$W$ channel), with similar performance for all benchmark samples.
In addition to satisfying a trigger requirement, every event is required to have a primary vertex with at least two tracks with $\pt > 500 $  \MeV~\cite{ATL-PHYS-PUB-2015-026}. At least one clean jet (which can be displaced) is required in the event for all channels.

The three analysis channels (CalR+2J, CalR+$W$, CalR+$Z$) exploit different final states but follow the same overall strategy: a per-jet NN provides discrimination  between signal-like jets, BIB-like jets and SM multijets. Additional per-event ML algorithms are trained to separate signal from remaining backgrounds at the event level. Some additional selections are applied to ensure consistent behaviour of the remaining population of events. The surviving events are then placed in a plane of two uncorrelated variables, enabling the data-driven ABCD method to be used for the final background estimate as described in Section~\ref{sec:bkgs}.

\subsection{CalR+2J channel selection}

The CalR+2J channel targets events containing one displaced jet, and two other jets that may be trackless.
Such a signature arises in the HS benchmark models where the mediator is produced by gluon--gluon fusion before decaying into a pair of LLPs. One LLP can decay in the hadronic part of the calorimeter (leading to a displaced jet where both the fermions are collimated into a single jet, with high probability to fulfil the criteria of the CalRatio triggers) and the other LLP can decay with a short enough decay length to produce two separate additional jets (one from each fermion). These jets would likely be reconstructed without associated tracks, and would be delayed relative to activity arriving from the interaction point in a straight line. Events containing mediators produced via vector-boson fusion would also enter this channel, although the additional information from the prompt jets would not be exploited. Hence, no sensitive additional contribution is expected from this production mode given its low cross-section compared with gluon--gluon fusion. A dedicated analysis would be needed to make the most of the vector boson fusion signature. The CalR+2J channel is complementary to the analysis in Ref.~\cite{EXOT-2019-23}, which was optimised for the case where both the LLPs led to merged displaced jets. In this analysis, the ``merged-resolved'' topology  (which can account for up to $27\%$ of events in the benchmark HS models) is explored for the first time.

Events entering this channel are required to satisfy a preselection consisting of having at least three clean jets and $\sum\Delta R_\textrm{min}$ being above 0.5, where the sum runs over jets with $\pt > 50$ \GeV.
This \pt threshold was chosen to be consistent with the previous analysis exploiting a similar topology~\cite{EXOT-2019-23}.
The jets with the three highest signal scores from the high-\et and low-\et NNs are selected, and one of these must be a CalRatio jet candidate.
The two other jets represent the resolved additional jets from a second LLP decay.
These jets and the jets with the three highest BIB scores must satisfy additional conditions. In practice the three jets with the highest signal and BIB scores can be the same jets.
The first condition is on jet time, which is the energy-weighted average time of the jet's constituent energy deposits, relative to the time it would take for a particle to travel at the speed of light directly from the interaction point to the jet's calorimeter location~\cite{PERF-2014-07}.
All jets should have a time within the range of $-3$ to $15$ ns.
This criterion aims to reject BIB jets (which do not need to pass via the interaction point and hence can arrive early) and noise-induced jet candidates (which are uniformly distributed in time), and preserve signal jets from LLPs moving relatively slowly with timings up to 15~ns.
Next, these jets should have $\logR > -1.5$.
Finally, the jets with the highest signal scores should not be in the barrel-endcap transition regions: $|\eta| \notin (1.45, 1.55)$.

\begin{figure}[t!]
\centering
\includegraphics[width=0.6\linewidth]{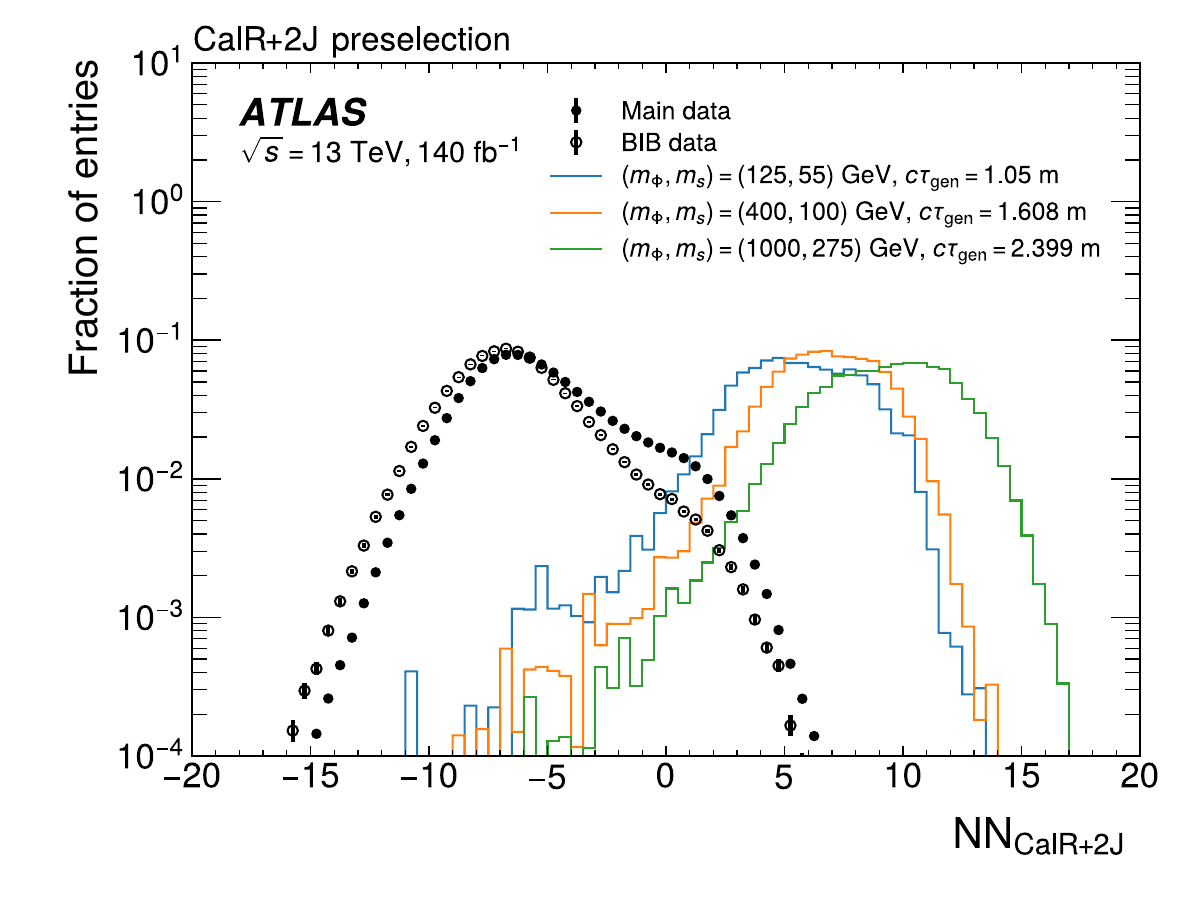}
\caption{Comparison of the NN$_\textrm{CalR+2J}$ distribution between signal events and background. Only signal events with at least one LLP decaying after the inner detector are shown. The main data, collected with the CalRatio triggers, are dominated by QCD multijet background.}
\label{fig:cr2j_nn_hist}
\end{figure}

An event-level NN (labelled NN$_{\textrm{CalR+2J}}$) is trained with the objective of removing all remaining BIB events from the selection, such that the only remaining background is multijet events. This facilitates the use of the ABCD method for the background estimate in Section~\ref{sec:bkgs}.
The input features for the NN$_{\textrm{CalR+2J}}$ includes jet-level and event-level variables. The jet observables include per-jet NN scores, kinematics, width and time. The event-level variables include the scalar sum hadronic energy of all jets, angular separations between jets, missing transverse energy and related quantities. The NN$_{\textrm{CalR+2J}}$ is trained on events from the BIB dataset and the HS signal samples after preselection.
It was trained to be decorrelated from the  $\sum\Delta R_\textrm{min}$ event-level variable by adding additional terms to the loss function, as described in Section~\ref{sec:disco}.
Figure~\ref{fig:cr2j_nn_hist} shows the distribution of the NN$_{\textrm{CalR+2J}}$ output for the BIB and main datasets and some example signal models.
BIB-like events are associated with very small values of the output.
Events in the BIB dataset with a NN$_{\textrm{CalR+2J}}$ output of three or higher do not show the characteristic timing and angular distributions of BIB and are indistinguishable from pure multijet events.
Thus events are required to have a NN$_{\textrm{CalR+2J}}$ score of at least three to be considered in this channel to effectively eliminate BIB-like events. Table~\ref{tab:CR2J_selections} lists the full selection criteria.

\begin{table}[]
\centering
\caption{The final selection criteria for the signal region of the CalR+2J channel. Region A is the signal region of the ABCD plane defined in Section~\ref{sec:bkgs} and includes all the other requirements listed in the table.}
\begin{tabular}{|r|l|l|l|}
\hline
Selection & CalR+2J \\
\hline
Trigger & Satisfy CalRatio trigger \\
Number of clean jets & $\geq 3$ \\
$\sumMinDR$ & $> 0.5$ \\
Trigger matching & At least one signal candidate \\
Signal/BIB jet candidate time & $-3 \textrm{ ns} < t < 15 \textrm{ ns}$ \\
Signal/BIB jet candidate $\log_{10} (E_\textrm{H} / E_\textrm{EM})$ & $> -1.5$ \\
Signal jet candidate $\eta$ & $\notin (1.45, 1.55)$ \\
$\textrm{NN}_\textrm{CalR+2J}$ & $\geq 3$ \\
\hline
\multirow{2}{*}{Region A} & $\sumMinDR \geq 0.71$ \\
& $\textrm{NN}_\textrm{CalR+2J} \geq 7.61$ \\
\hline
\end{tabular}

\label{tab:CR2J_selections}
\end{table}

\subsection{CalR+$W$ and CalR+$Z$ channel selections}

\subsubsection{Preselection}
\label{sec_cr_1or2l}

The leptonic channels share a common preselection. At least one trackless jet  with $\pt > 40 \gev$ is required, and the $\sum\Delta R_\textrm{min}$ is required to be above 0.5, where the sum runs over jets with $\pt > 40$ \GeV.
In both the channels, the jet with the highest low-$E_\textrm{T}$ NN signal-score ($j^{\textrm{sig} 1\ell}$) is required to be trackless and to have $\pt > 50$ \GeV, $\lograt> -1$, a time in the range of $(-3, 15)$~ns and not be in the transition regions between the barrel and endcap.
The NN signal-score of this jet is required to be above 0.4.
Both channels use the low-$E_\textrm{T}$ NN scores to rank jets from most signal-like to least signal-like.
As it was already observed in Ref.~\cite{EXOT-2019-23}, even for high-$E_\textrm{T}$ input signal samples the low-$E_\textrm{T}$ NN scores provide a better separation from background in cases where the LLP boost is moderate.

\subsubsection{CalR+$W$ channel selections}
\label{sec_cr_1l}

The CalR+$W$ channel targets scenarios where the HS mediator is produced in association with a $W$ boson, where the $W$ boson decays leptonically. While this production mode has a lower expected cross-section than gluon--gluon fusion, the additional lepton from the $W$ boson decay provides a clear signature on which to trigger. It also renders most NCBs negligible, and makes it possible to require only one displaced jet while two were required in Ref.~\cite{EXOT-2019-23}. In addition to the HS models, ALP models also predict events in this topology.

After the application of the preselection, a $W$ boson candidate is required in the event. Events with more than one lepton are rejected, to maintain orthogonality with the CalR+$Z$ channel. The most signal-like jet must be separated in the azimuthal direction from the \met: $\Delta \phi (j^{\textrm{sig} 1\ell}, \met)>0.5$. %
To be orthogonal to the CalR+$Z$ channel, events with more than one baseline lepton are vetoed.
Finally, the CalR+$W$ channel exploits a set of Boosted Decision Trees (BDTs) as discriminants. These are trained with XGBoost~\cite{Chen:2016:XST:2939672.2939785} with the objective of discriminating signal from background events after preselection and  $W$ boson identification is applied. Three BDTs are trained corresponding to three input signal sample sets. The first uses a combination of all $W$ALP mass points, in which there is only one LLP in the event. The other two BDTs are trained on low ($m_{\Phi}\leq 200$ \GeV) and high ( $m_{\Phi} > 200$ \GeV)
HS mediator mass events with an associated $W$ boson. In all cases, the BDT is trained to discriminate the signal against MC simulated events from all background processes listed in Section~\ref{sec:datasets} for this channel. These three BDTs are hereafter referred to as BDT$_{\textrm{CalR+}W}^{\textrm{ALP}}$,  BDT$_{\textrm{CalR+}W}^{\textrm{low-}\ET}$ and BDT$_{\textrm{CalR+}W}^{\textrm{high-}\ET}$.
For signal, only events with an odd event number are used for the training, leaving the even numbered events for evaluation. The  input variables include information from jets (such as \pt, jet width and per-jet NN scores), leptons (such as \pt and angles between leptons and signal-like jets) and event-level observables (such as missing transverse energy in the event, reconstructed vector boson kinematics).

\begin{figure}[t]
\centering
\subfloat[]{
\includegraphics[width=0.4\linewidth]{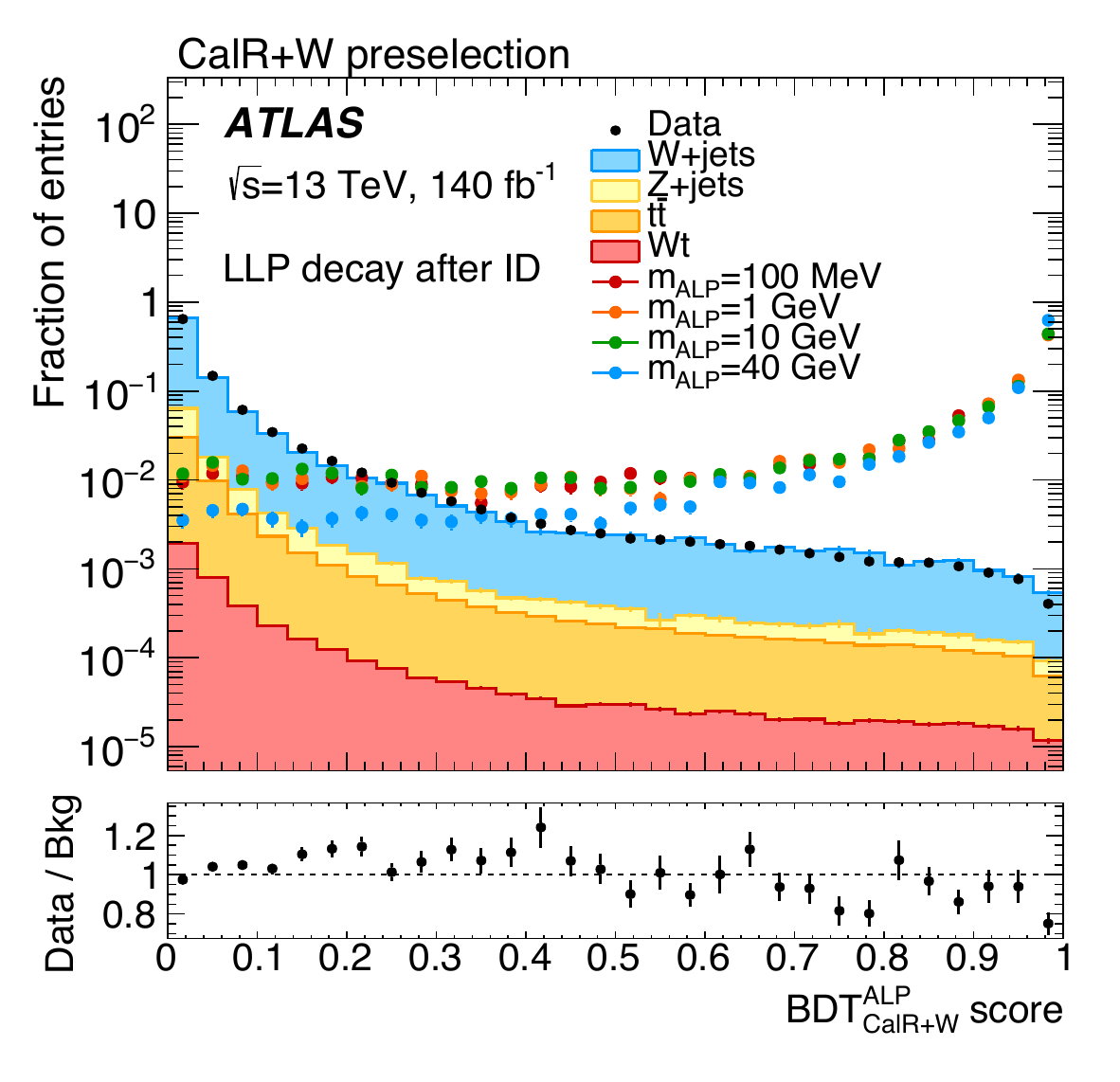}
\label{fig:CalRatioW_BDT_alp}}\\
\subfloat[]{\includegraphics[width=0.4\linewidth]{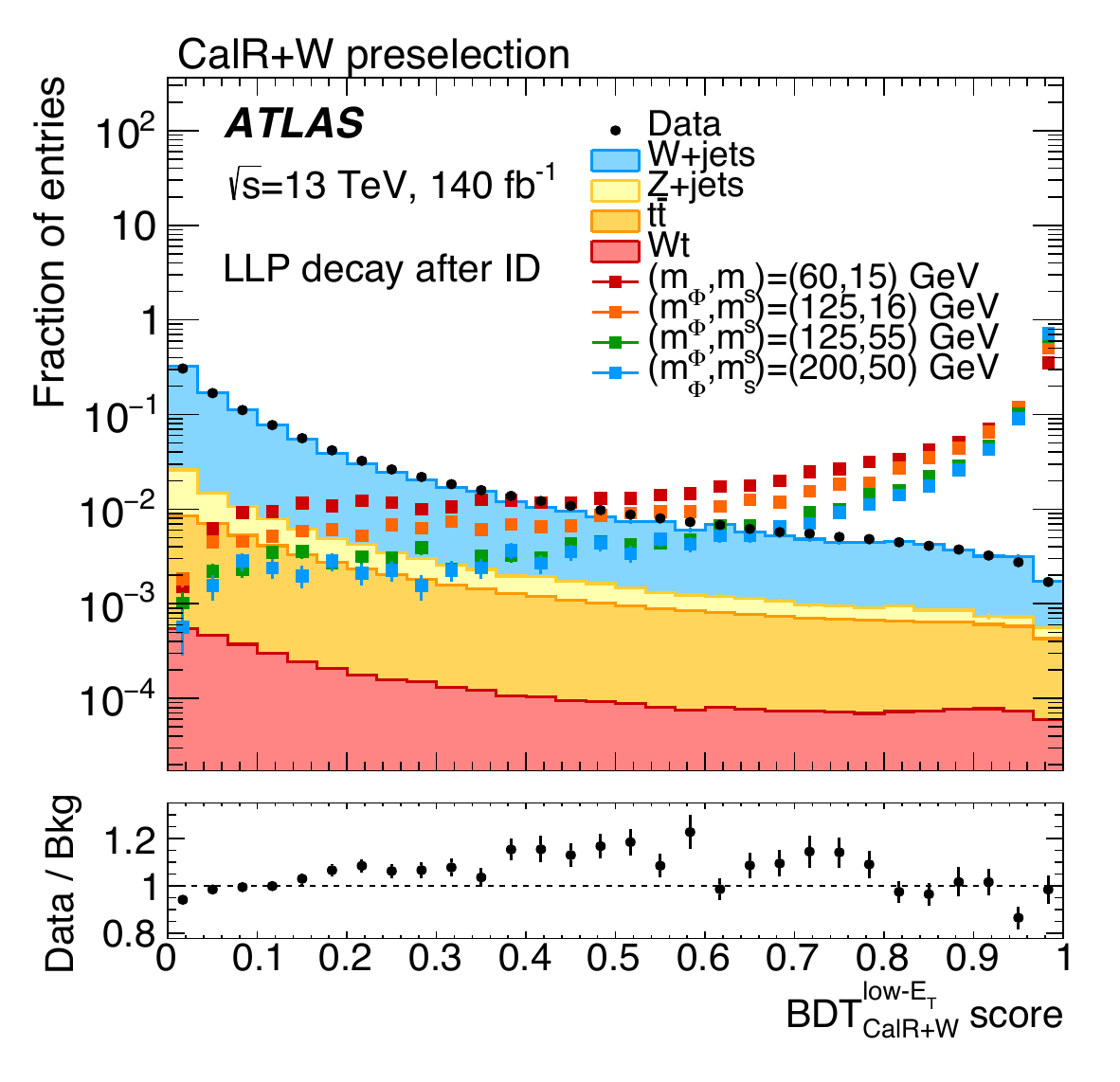}
\label{fig:CalRatioW_BDT_lo}}
\subfloat[]{
\includegraphics[width=0.4\linewidth]{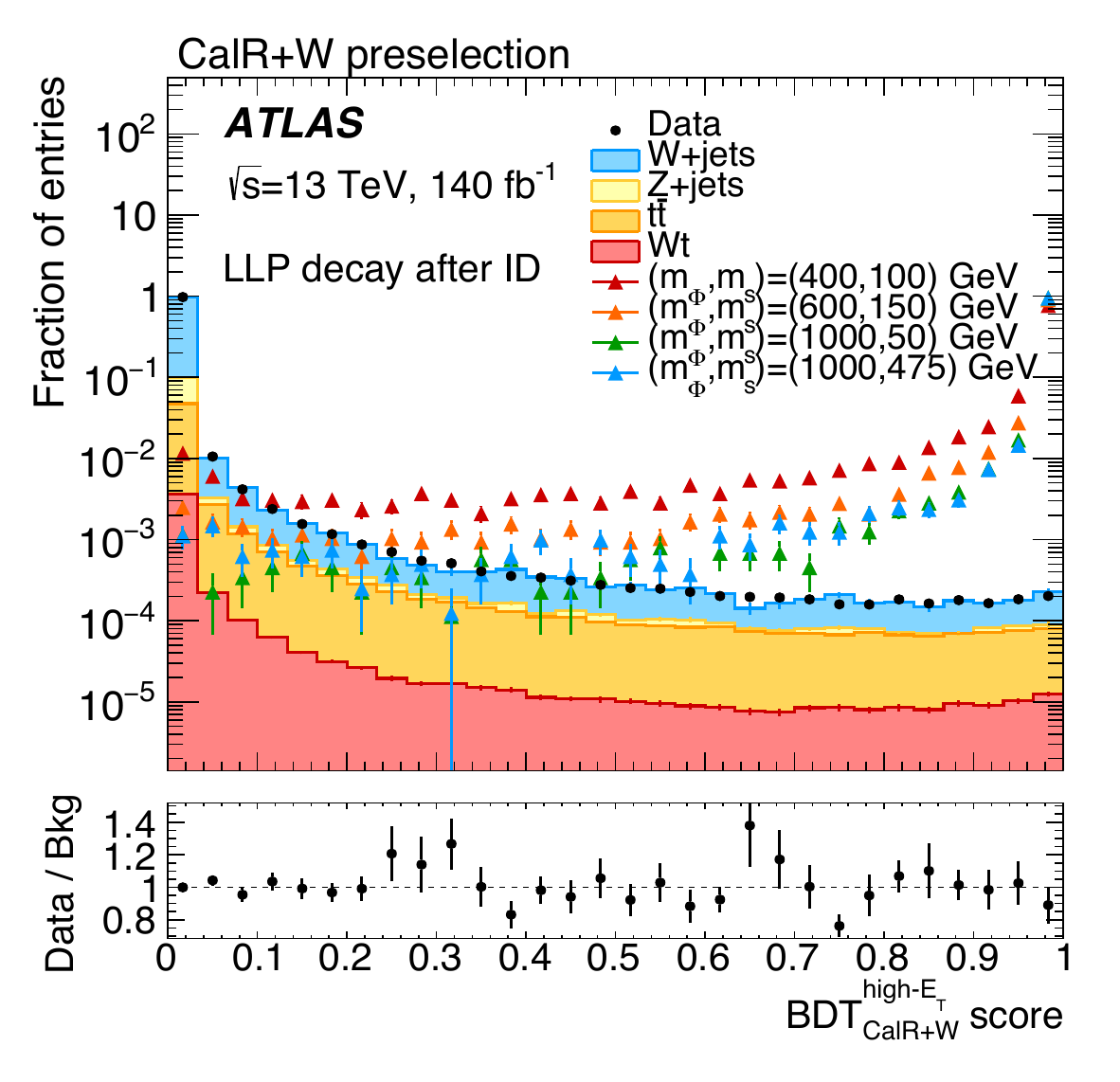}
\label{fig:CalRatioW_BDT_hi}}

\caption{Comparison of the distribution of (a)  BDT$_{\textrm{CalR+}W}^{\textrm{ALP}}$, (b) BDT$_{\textrm{CalR+}W}^{\textrm{low-}\ET}$, (c) BDT$_{\textrm{CalR+}W}^{\textrm{high-}\ET}$ scores between signal events and background, after preselection. In all cases, only signal events with at least one LLP decaying after the ID are shown. The simulated background events are shown only for comparison at the preselection stage, and are not used in the final background estimate.}
\label{fig:BDTsW}
\end{figure}

Figures~\ref{fig:CalRatioW_BDT_alp}, \ref{fig:CalRatioW_BDT_lo} and~\ref{fig:CalRatioW_BDT_hi} show the distribution of BDT$_{\textrm{CalR+}W}^{\textrm{ALP}}$,  BDT$_{\textrm{CalR+}W}^{\textrm{low-}\ET}$ and BDT$_{\textrm{CalR+}W}^{\textrm{high-}\ET}$ on four signal models, data and background MC\@.
The simulated background events are shown only for comparison at the preselection stage, and are not used in the final background estimate. Dedicated uncertainties are derived to cover the effect of mis-modelling of input variables on signal efficiency, as detailed in Section~\ref{sec:systematics}.
Three sets of selection criteria are defined, each exploiting one of these BDTs. The selection criteria are listed in Table~\ref{tab:CRW_selections}.

\begin{table}[t]
\centering
\caption{The final selection criteria for the signal regions of the CalR+$W$ channel. Region A is the signal region of the ABCD plane defined in Section~\ref{sec:bkgs} and includes all the other requirements listed in the table.}
\begin{tabular}{|r|l|l|l|}
\hline
Selection &  CalR+$W$ $WA$LP  & CalR+$W$  low-\ET   &  CalR+$W$  high-\ET   \\
\hline
Vector boson candidates                         & 0 $Z$, 1 $W$       & 0 $Z$, 1 $W$      & 0 $Z$, 1 $W$   \\

BDT score                                & BDT$_{\textrm{CalR+}W}^{\textrm{ALP}}$  $> 0.82$                &  BDT$_{\textrm{CalR+}W}^{\textrm{low-}\ET}> 0.92$          & BDT$_{\textrm{CalR+}W}^{\textrm{high-}\ET}$  $> 0.89$    \\
$j^{\textrm{sig} 1\ell}$ \lograt                     & $>1$                                  &  $>1$                                 & --                                 \\
$j^{\textrm{sig} 1\ell}$ \pt                          & $>70$ \GeV                             &  $>60$ \GeV                            &  $>100$ \GeV                         \\
Lepton \pt                                &  --                                  &  $> 40$ \GeV                           & $ > 60$ \GeV                        \\
$\Delta\phi (\text{lepton, }E_T^\mathrm{miss})$   & $ <1.5$                               &  --                                  &  --                                \\
\hline
\multirow{2}{*}{Region A}                 & BDT$_{\textrm{CalR+}W}^{\textrm{ALP}} \geq 0.975$               & BDT$_{\textrm{CalR+}W}^{\textrm{low-}\ET}\geq 0.985$       & BDT$_{\textrm{CalR+}W}^{\textrm{high-}\ET} \geq 0.99$    \\
&  $\sum\Delta R_\textrm{min}\geq 1.1$  &  $\sum\Delta R_\textrm{min}\geq
1.4$  & $\sum\Delta R_\textrm{min}\geq 1.1$ \\
\hline
\end{tabular}

\label{tab:CRW_selections}
\end{table}

\begin{figure}[t]
\centering
\subfloat[]{
\includegraphics[width=0.4\linewidth]{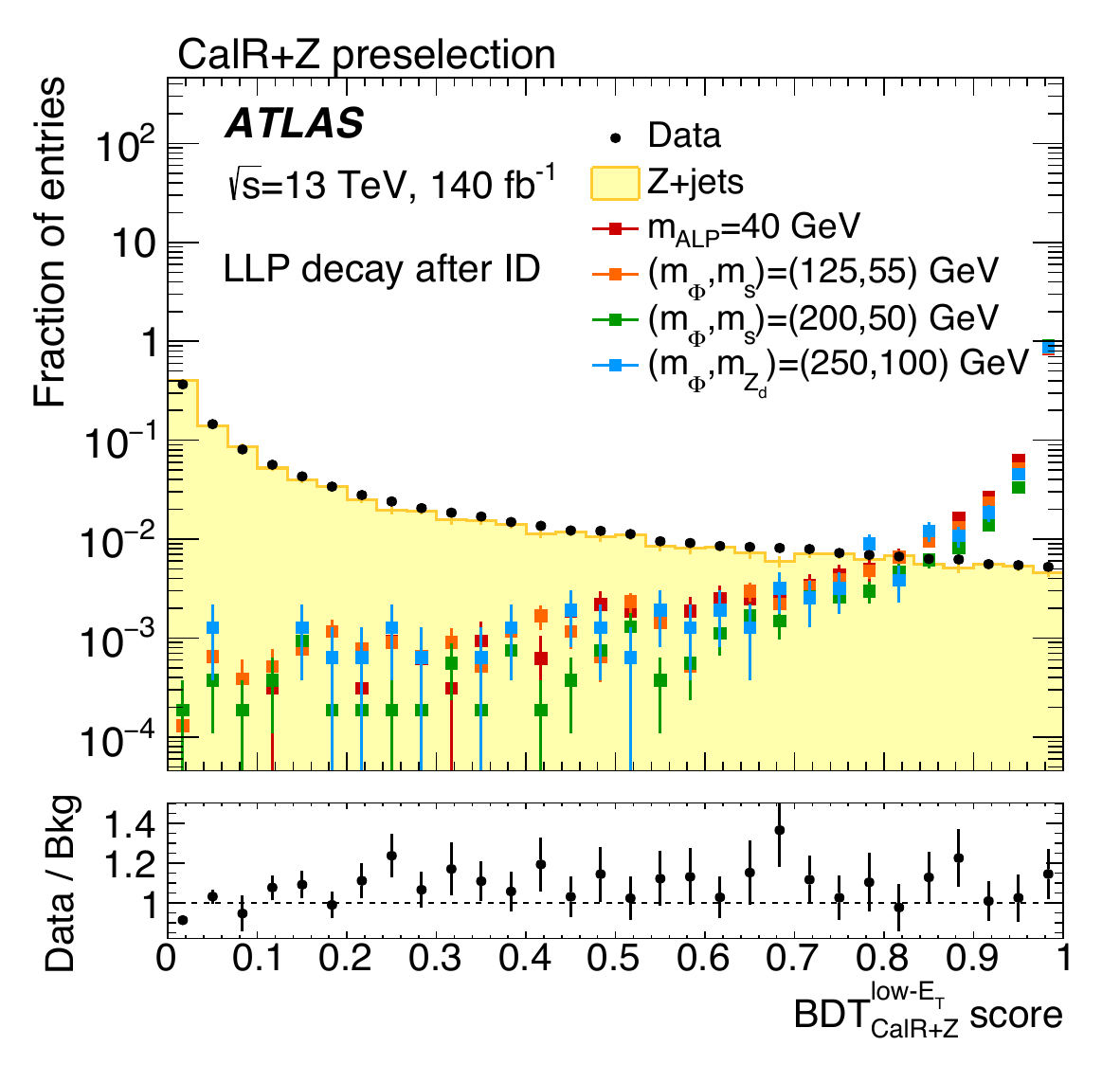}
\label{fig:CalRatioZ_BDT_lo}}
\subfloat[]{
\includegraphics[width=0.4\linewidth]{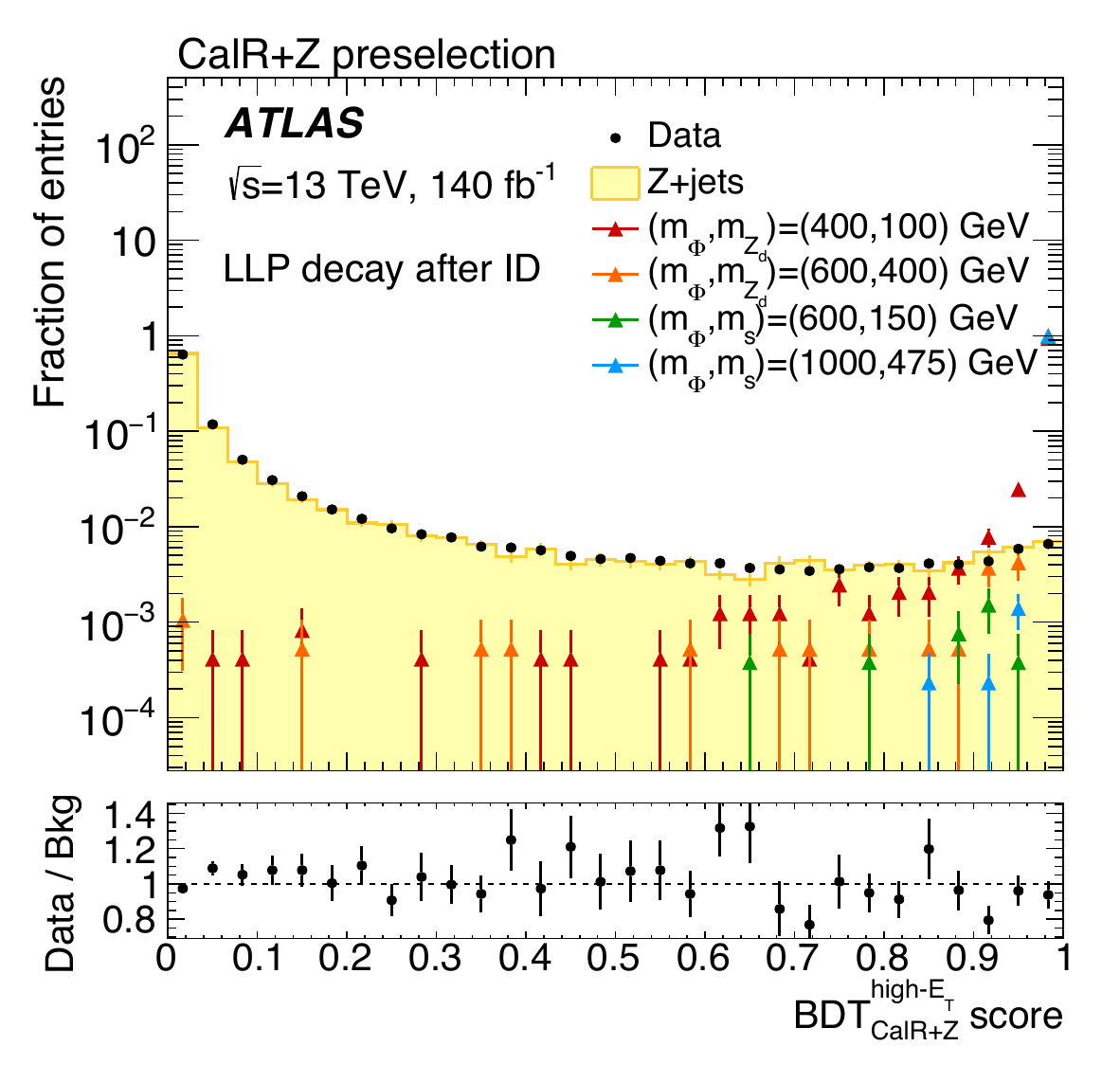}
\label{fig:CalRatioZ_BDT_hi}
}
\caption{Comparison of the distribution of (a) BDT$_{\textrm{CalR+}Z}^{\textrm{low-}\ET}$ and (b) BDT$_{\textrm{CalR+}Z}^{\textrm{high-}\ET}$ scores between signal events and background, after preselection. In all cases, only signal events with at least one LLP decaying after the ID are shown. The simulated background events are shown only for comparison at the preselection stage, and are not used in the final background estimate.}
\label{fig:BDTsZ}
\end{figure}

\subsubsection{CalR+$Z$ channel selection}
\label{sec_cr_2l}

The CalR+$Z$ channel targets the topology where the LLP is accompanied by a $Z$ boson decaying into two charged leptons. The motivation is identical to that described in Section~\ref{sec_cr_1l}.
In addition to the benchmark HS and ALP models, this channel considers the study of the long-lived dark photon ($Z_d$) model described in Section~\ref{sec:datasets}.

After passing the preselection, events entering the CalR+$Z$ channel must contain a single $Z$ boson candidate. Two different BDTs are trained with XGBoost to discriminate signal from background events. The resulting BDT scores are used for a data-driven background estimate (see Section \ref{sec:bkgs} for details). Each of these BDTs aims at an optimal discrimination in a specific range of masses:
the BDT$_{\textrm{CalR+}Z}^{\textrm{low-}\ET}$ is trained with a combination of signal samples including all the $Z$ALP, Z${_d}$ and $Z$HS model with masses of the mediator $m_{\Phi} \leq 250$ \GeV; the BDT$_{\textrm{CalR+}Z}^{\textrm{high-}\ET}$ is trained with Z${_d}$ and $Z$+HS samples with $m_{\Phi} > 250$ \GeV. In the CalR+$Z$ channel, unlike the CalR+$W$ channel, a dedicated selection targeting ALP models was not needed: this is because the reconstruction of the associated vector boson is more straightforward, permitting a higher signal purity even when the ALP models are targeted together with others using the BDT$_{\textrm{CalR+}Z}^{\textrm{low-}\ET}$. The BDTs are trained to discriminate the signal against the background processes listed in Section~\ref{sec:datasets} for this channel. The input variables are similar to those of the CalR+$W$ channel BDTs.
Figures~\ref{fig:CalRatioZ_BDT_lo} and~\ref{fig:CalRatioZ_BDT_hi} show the distributions of the BDT$_{\textrm{CalR+}Z}^{\textrm{low-}\ET}$ and BDT$_{\textrm{CalR+}Z}^{\textrm{high-}\ET}$ on several signal models, on data and background MC\@. Two selection criteria are defined, one for each BDT, as outlined in Table~\ref{tab:CRZ_selections}.

\begin{table}[]
\centering
\caption{The final selection criteria for the signal regions of the CalR+$Z$ channel. Region A is the signal region of the ABCD plane defined in Section~\ref{sec:bkgs} and includes all the other requirements listed in the table.}
\begin{tabular}{|r|l|l|}
\hline
Selection &  CalR+$Z$  low-\ET & CalR+$Z$  high-\ET \\
\hline
Vector boson candidates                         & 1 $Z$, 0 $W$        & 1 $Z$, 0 $W$  \\
BDT score                                & BDT$_{\textrm{CalR+}Z}^{\textrm{low-}\ET}$ score $> 0.6$         &  BDT$_{\textrm{CalR+}Z}^{\textrm{high-}\ET}$ score $> 0.7$        \\
$j^{\textrm{sig} 1\ell}$ \lograt                     & $>0.8$                                &   $>0.8$                              \\
$j^{\textrm{sig} 1\ell}$ \pt                          &$>80$ \GeV                              &  $>70$ \GeV                            \\
Lepton \pt                                &  $> 70$ \GeV                           &          $> 60$ \GeV                   \\
\hline
\multirow{2}{*}{Region A}                 & BDT$_{\textrm{CalR+}Z}^{\textrm{low-}\ET}$ score $> 0.99 $       & BDT$_{\textrm{CalR+}Z}^{\textrm{high-}\ET}$ score $> 0.985$      \\
&  $\sum\Delta R_\textrm{min}\geq 0.9$  &  $\sum\Delta R_\textrm{min}\geq 1$   \\

\hline
\end{tabular}

\label{tab:CRZ_selections}
\end{table}

\FloatBarrier


\section{Background estimation}
\label{sec:bkgs}

Following the application of the selections, the contributions from cosmic rays and all subdominant SM processes are found to be negligible. Some events from the BIB dataset still pass the selections, but these remaining events show the same behaviour as multijet events. This leaves only one population of background events in each channel, namely multijet events in the CalR+2J channel and $W$/$Z$+jets in the CalR+$W$ and CalR+$Z$ channels, respectively. The CalR+$W$ channel contains residual traces of $t\bar{t}$ events, but these show the same behaviour as the dominant background and are therefore grouped as the same population. As these background sources are not well modelled, a data-driven method is used to estimate the background yields in the final signal regions.
The data-driven ABCD method is employed to estimate the contribution from the dominant background in each of the final selections. The ABCD method is described extensively in other publications, such as Ref.~\cite{EXOT-2019-23}; the main points are summarised here. In a plane defined by two uncorrelated variables, four regions A, B, C, and D are defined using horizontal and vertical boundaries. When there is one predominant background population (typically concentrated in regions B, C and D) and a signal population primarily in region A, the background yield in region A ($\text{N}_\text{A}$) can be calculated from the yields in the other regions ($\text{N}_\text{B}$, $\text{N}_\text{C}$ and $\text{N}_\text{D}$) using the formula $\text{N}_\text{A} = (\text{N}_\text{B} \times \text{N}_\text{C}) / \text{N}_\text{D}$. Leakage of signal events into the other regions can be accounted for using a simultaneous fit. All three channels exploit ABCD planes for their background estimates, although the axes are defined differently in each case.

\subsection{CalR+2J final selection and the ABCDisCo method}
\label{sec:disco}

\begin{figure}[t]
\centering
\subfloat[]{
\includegraphics[width=0.49\linewidth]{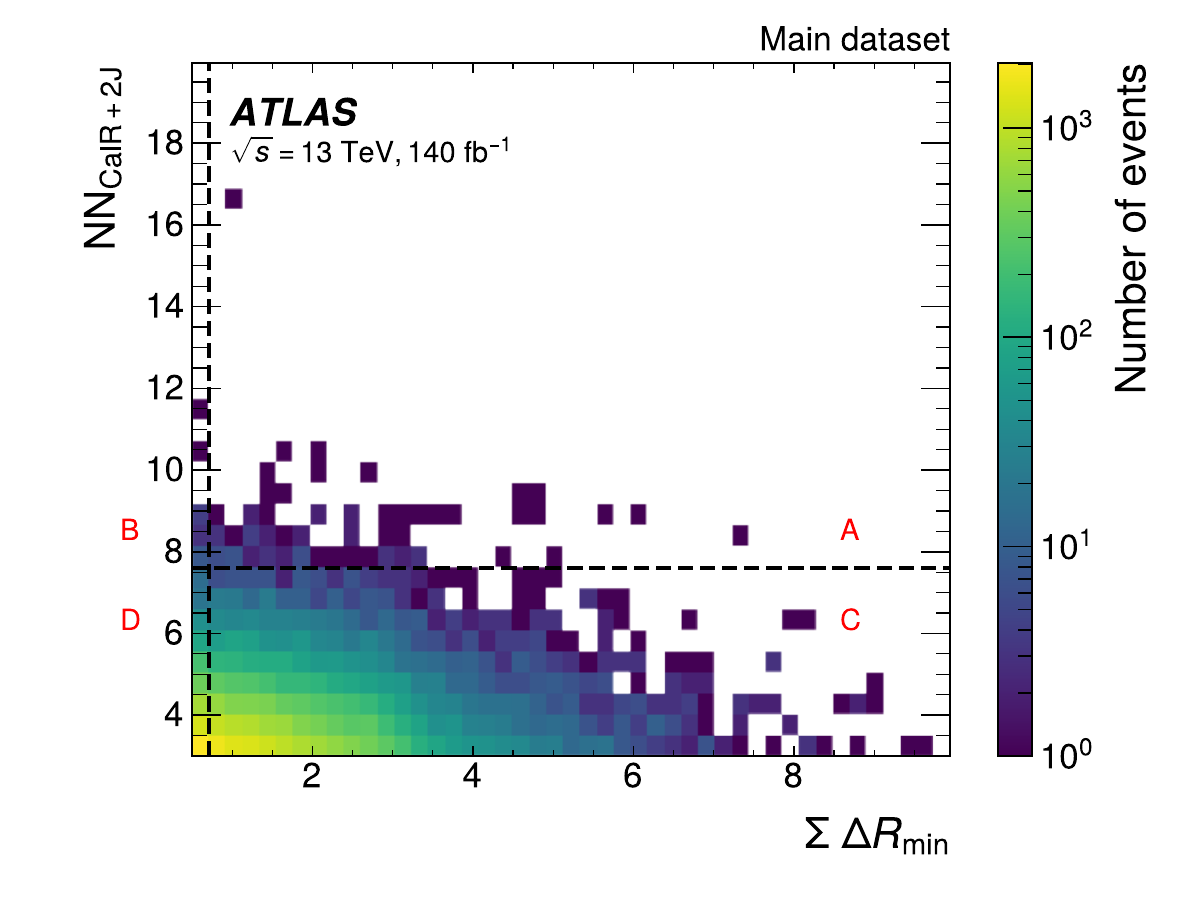}
}
\subfloat[]{
\includegraphics[width=0.49\linewidth]{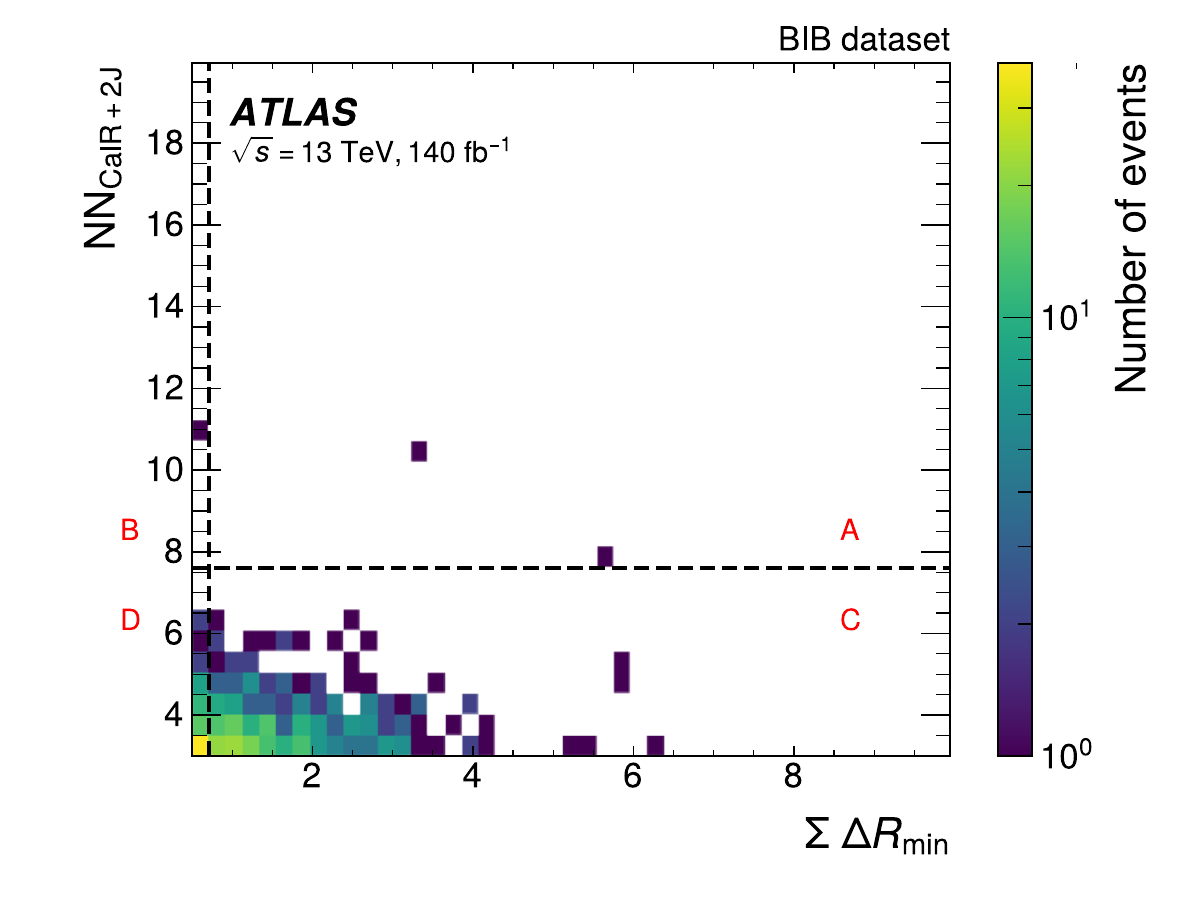}
}\\
\subfloat[]{
\includegraphics[width=0.49\linewidth]{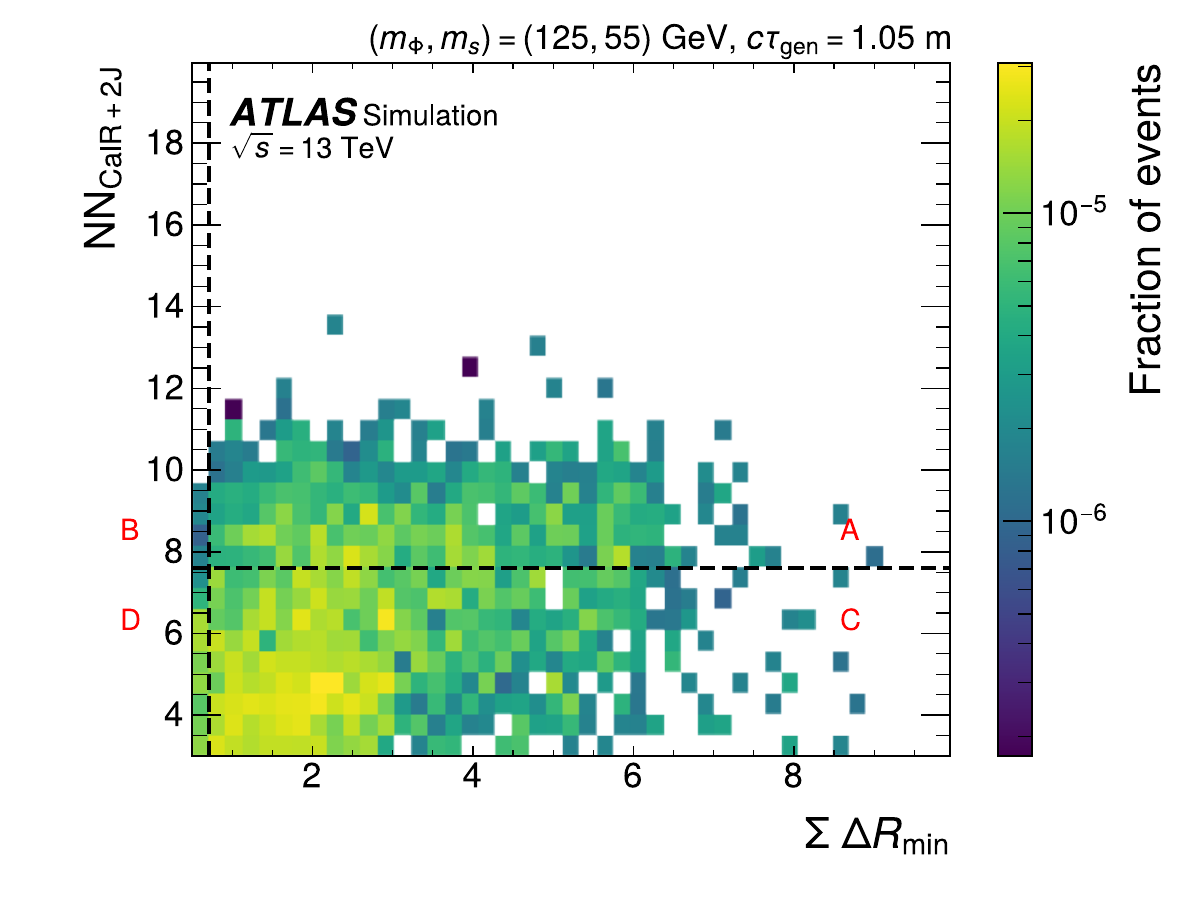}
}
\subfloat[]{
\includegraphics[width=0.49\linewidth]{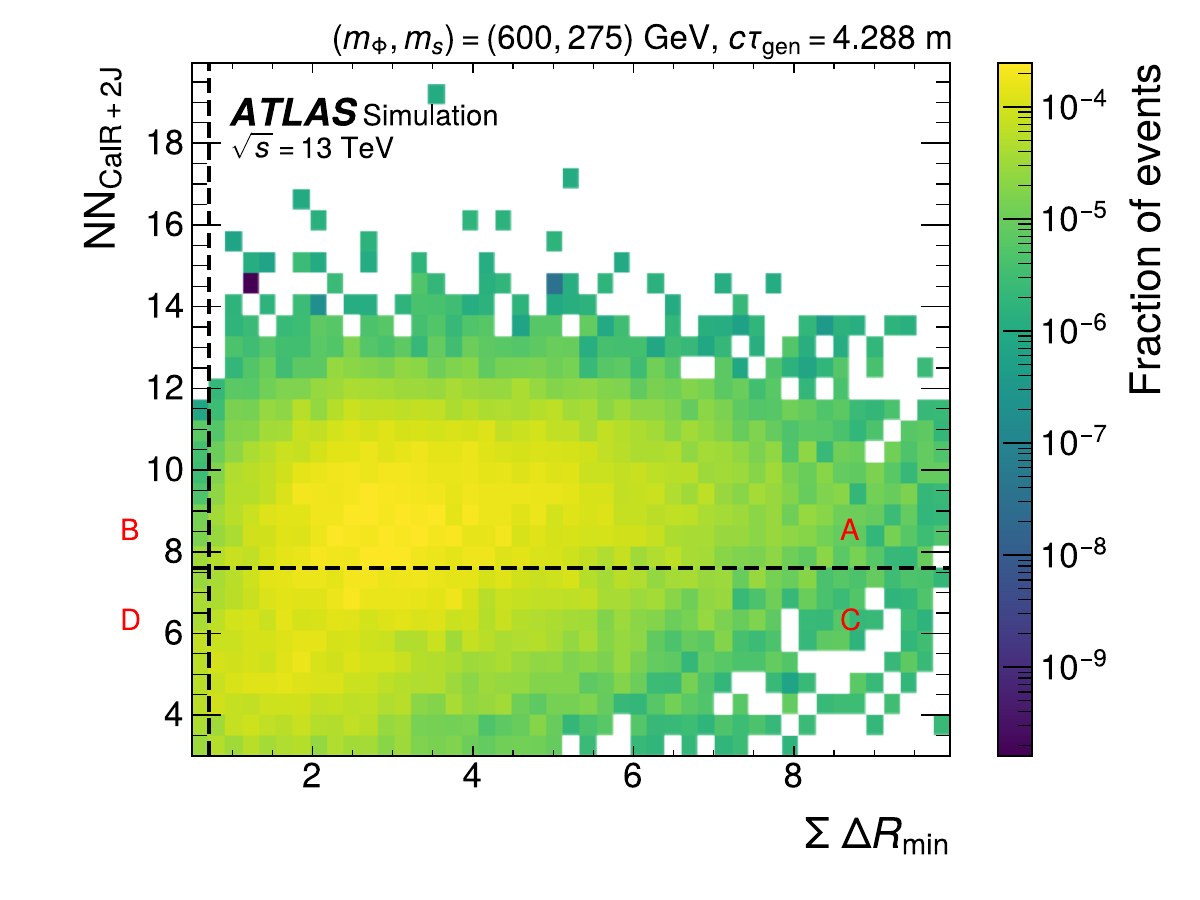}
}
\caption{Distribution of events in the two-dimensional plane of NN$_
\textrm{CalR+2J}$ vs. \sumMinDR ~in the CalR+2J channel for (a) the main dataset, (b) the BIB datasets, and (c, d) example benchmark signal samples. Final regions A, B, C and D are marked in red.
}
\label{fig:cr2j_nn_plane_plots}
\end{figure}

The ABCD plane axes are usually chosen from the list of available variables that separate signal and background well.
In cases where it is difficult to find two uncorrelated variables with enough separation power, another approach is to train a ML discriminant, which is required to be uncorrelated to another variable by construction.
The CalR+2J channel is such a case: unlike in Ref.~\cite{EXOT-2019-23}, the $\sum\Delta R_\textrm{min}$ becomes visibly correlated with displaced jet kinematic information as additional jets from LLP decays are included. A method called ABCDisCo uses the \textit{dis}tance \textit{co}rrelation (DisCo) between the ML discriminant and another event-level variable, and involves exploiting terms to the loss function of a neural network that penalises the network if these have a large correlation~\cite{Kasieczka:2020pil}.
The ABCDisCo method was employed in the CalR+2J channel to build an ABCD plane constructed from \sumMinDR\ and the NN output.
This is the same NN that is used to eliminate BIB-like events.
Even though this NN was only trained with the BIB dataset as background, it has a good discrimination power between signal and multijet events, which are also present in the BIB dataset.
The Pearson correlation coefficient for the resulting pair of discriminants is 0.04, small by construction.
The final region A is defined by the requirements that $\sumMinDR \geq 0.71$ and  NN$_\textrm{CalR+2J} \geq 7.61$. Regions B, C and D are obtained by inverting one or both the requirements, as shown in Figure~\ref{fig:cr2j_nn_plane_plots} for the main CalRatio dataset, the BIB dataset and two of the signal samples.

Regions B, C, and D were used to create the validation regions VR$_\textrm{BD}$ and VR$_\textrm{CD}$.
VR$_\textrm{BD}$ consists of the nominal B and D regions combined and is divided into four alternative ABCD regions.
The ABCD estimate procedure was tested with these regions and with variations on the cut values in each ABCD plane variable.
The predictions are found to be in agreement with the observed number of events within the statistical uncertainties.
The same procedure was performed for VR$_\textrm{CD}$, composed of the nominal C and D regions.
In VR$_\textrm{CD}$, the observed number of events was consistently about 5\% higher than estimated by the ABCD method, which was taken as the systematic uncertainty in the final ABCD prediction in this plane, as discussed in Section~\ref{sec:systematics}.

\subsection{CalR+$W$ and CalR+$Z$ final selections and background estimate}
\label{ABCD_validation_CRW}

The standard ABCD method is used to estimate the remaining background contribution in the five signal regions of the CalR+$W$/$Z$ channels, with the final ABCD planes defined by $\sum\Delta R_\textrm{min}$ and the BDT score relevant to each region. These variables are found to be uncorrelated after preselection, with Pearson correlation coefficients between 0.003 and 0.017 in absolute value in all cases. The ABCDisCo method is not needed in this case because the $\sum\Delta R_\textrm{min}$ is found to be sufficiently well decorrelated to the BDT scores (since the targeted LLP decays always happen outside the tracker).
The final ABCD boundaries are set through a selection optimisation procedure performed together with several additional event-level variables (such as kinematic information from the vector boson candidate, angular separations between the jets and leptons, and invariant masses of the possible lepton-jet combinations) to obtain the final selections that maximize sensitivity while reducing the signal leakage in regions B, C and D and keeping the statistical error on the estimated number of events in A below $35\%$. These selections, together with the definition of region A, are shown in Tables~\ref{tab:CRW_selections} and~\ref{tab:CRZ_selections}. The definitions of regions B, C and D can be obtained by inverting one or both the requirements that define region A.

The ABCD method is tested in validation regions that are orthogonal to the signal regions. In the CalR+$W$ channel, validation regions are defined in regions C and D of the main ABCD plane (VR$_\textrm{CD}$), with upper BDT score boundary as close as possible to region A but ensuring a low signal contamination.
In the CalR+$Z$ channel, part of regions C and D of the primary ABCD planes are used to define the validation regions. A looser set of criteria was employed for a dedicated validation region (VR$_\textrm{CD}^\textrm{Loose}$) to mitigate the significant statistical uncertainty observed in the original region due to limited number of events.
In all channels and all validation regions, the agreement between the expected and observed number of events is found to be within statistical uncertainty, and for a wide variety of thresholds on each variable.

\begin{figure}[t]
\centering
\subfloat[]{
\includegraphics[width=0.32\linewidth]{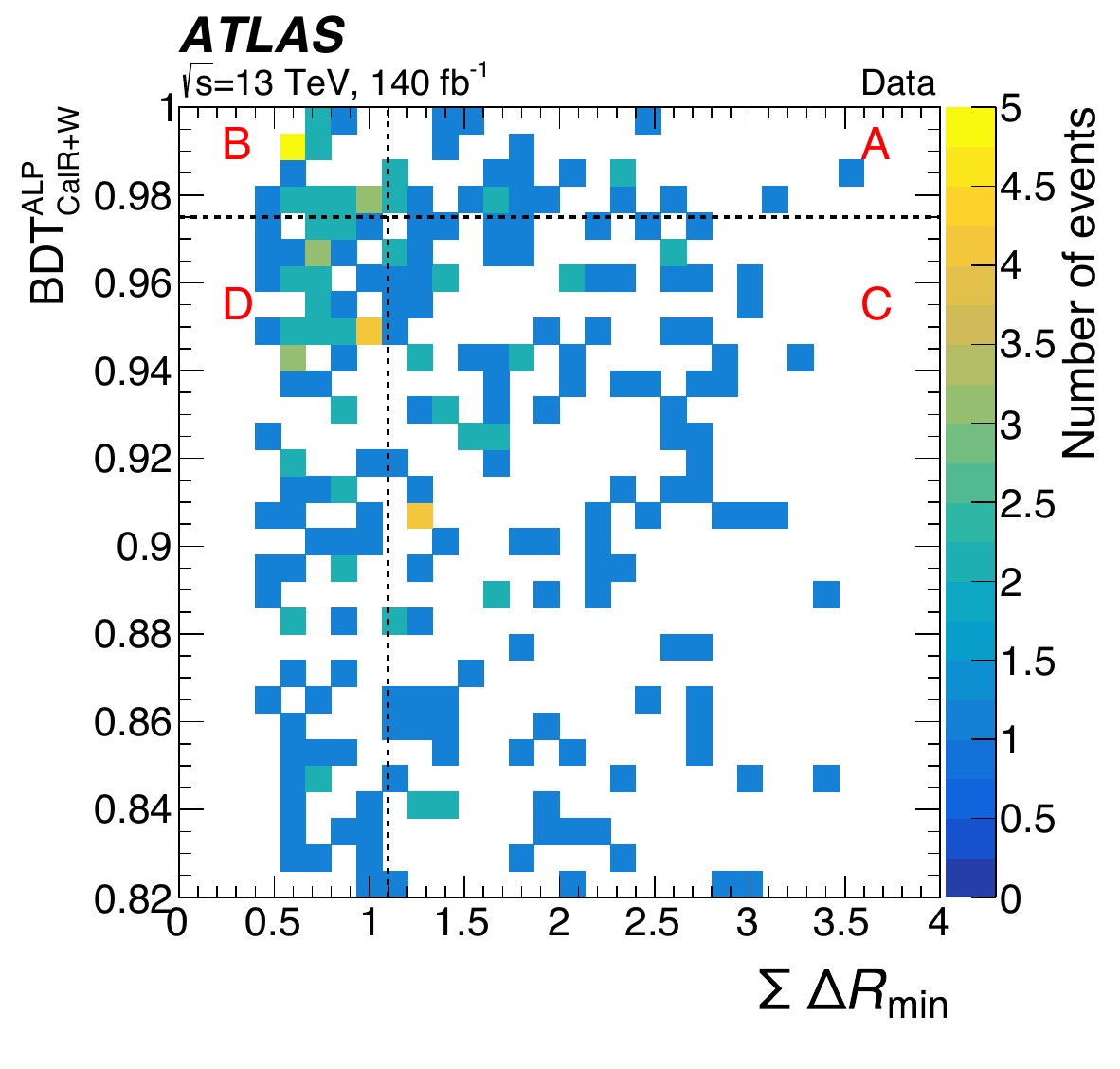}
} \subfloat[]{
\includegraphics[width=0.32\linewidth]{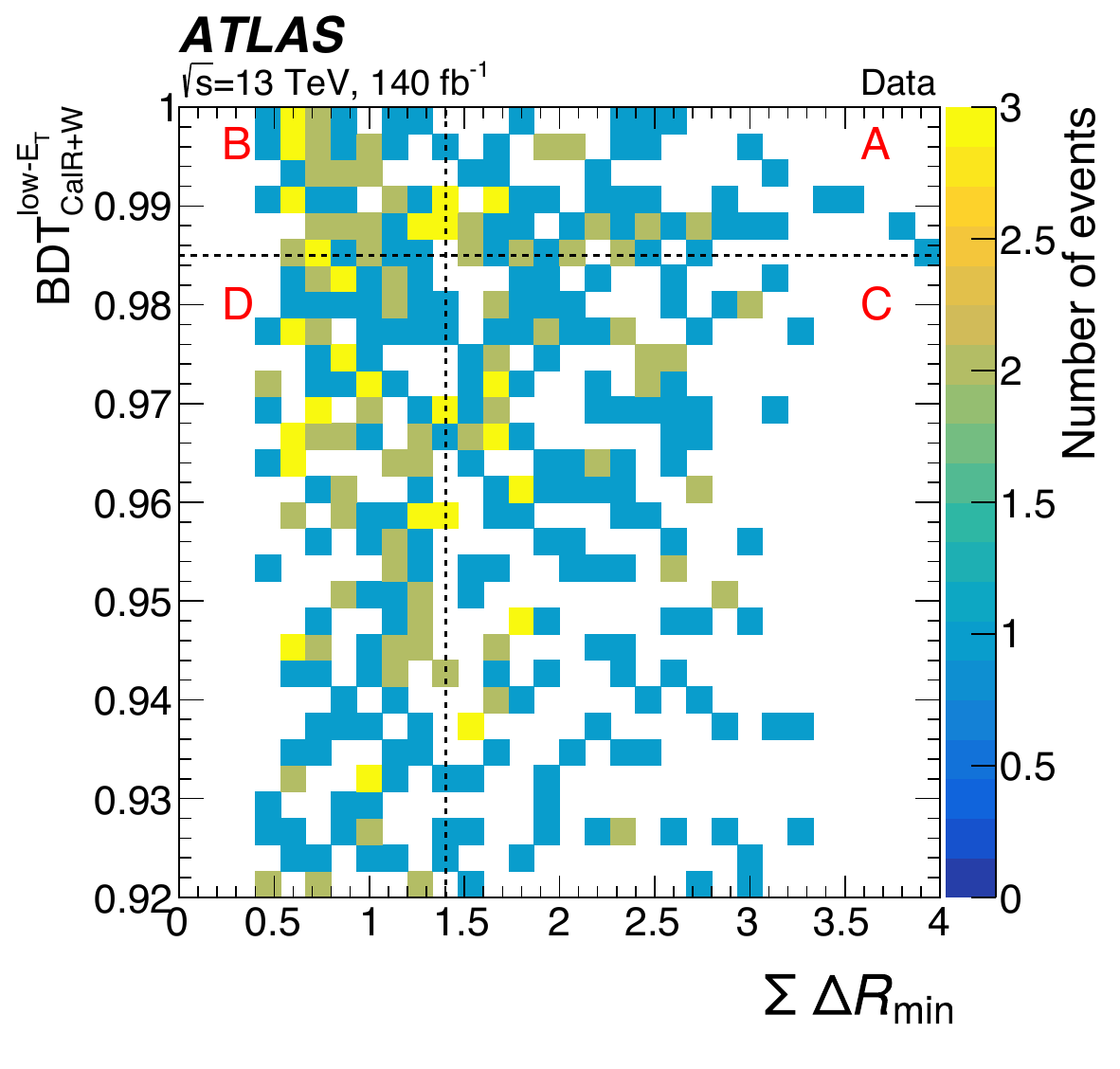}
} \subfloat[]{
\includegraphics[width=0.32\linewidth]{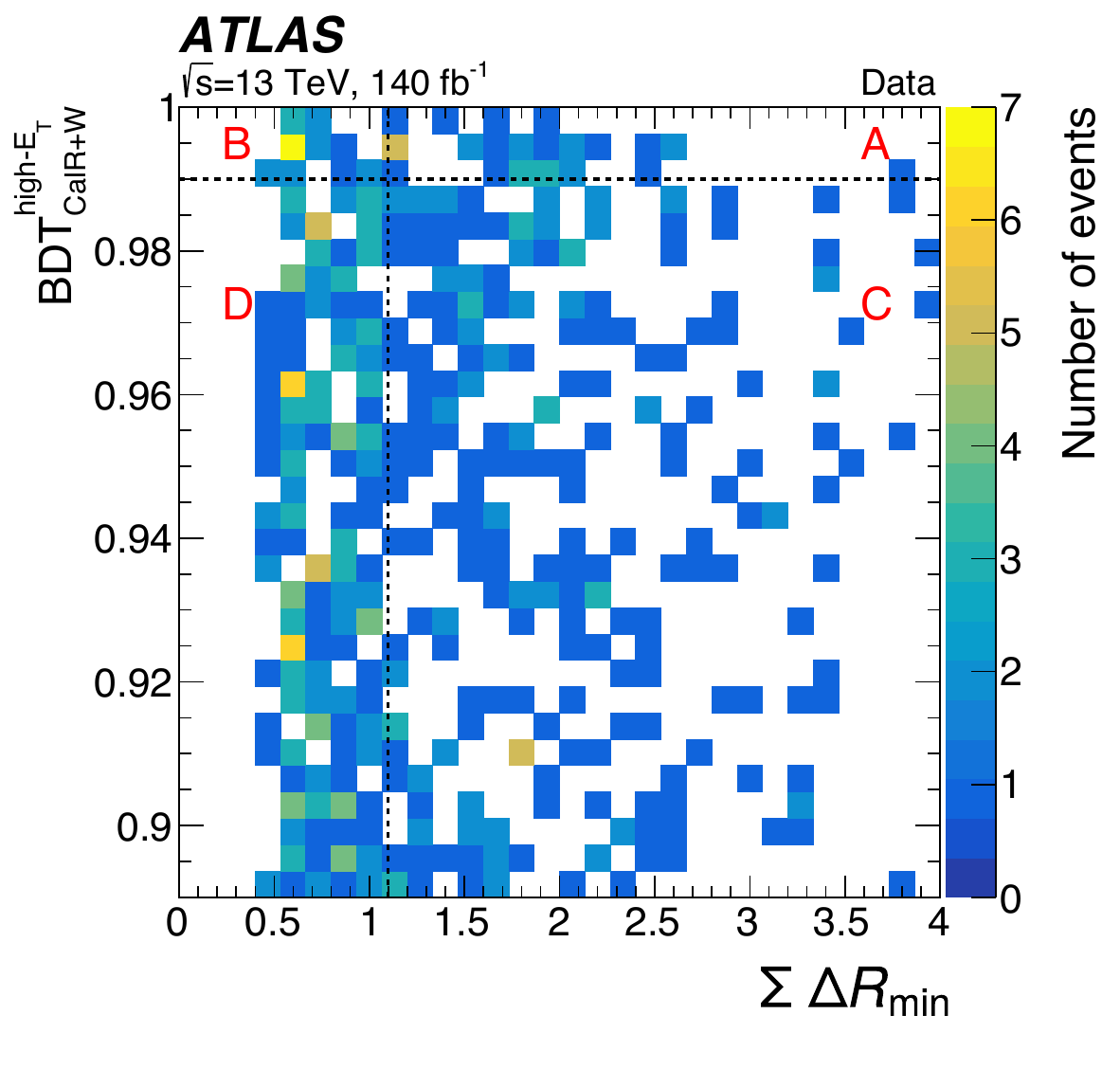}}\\
\subfloat[]{
\includegraphics[width=0.32\linewidth]{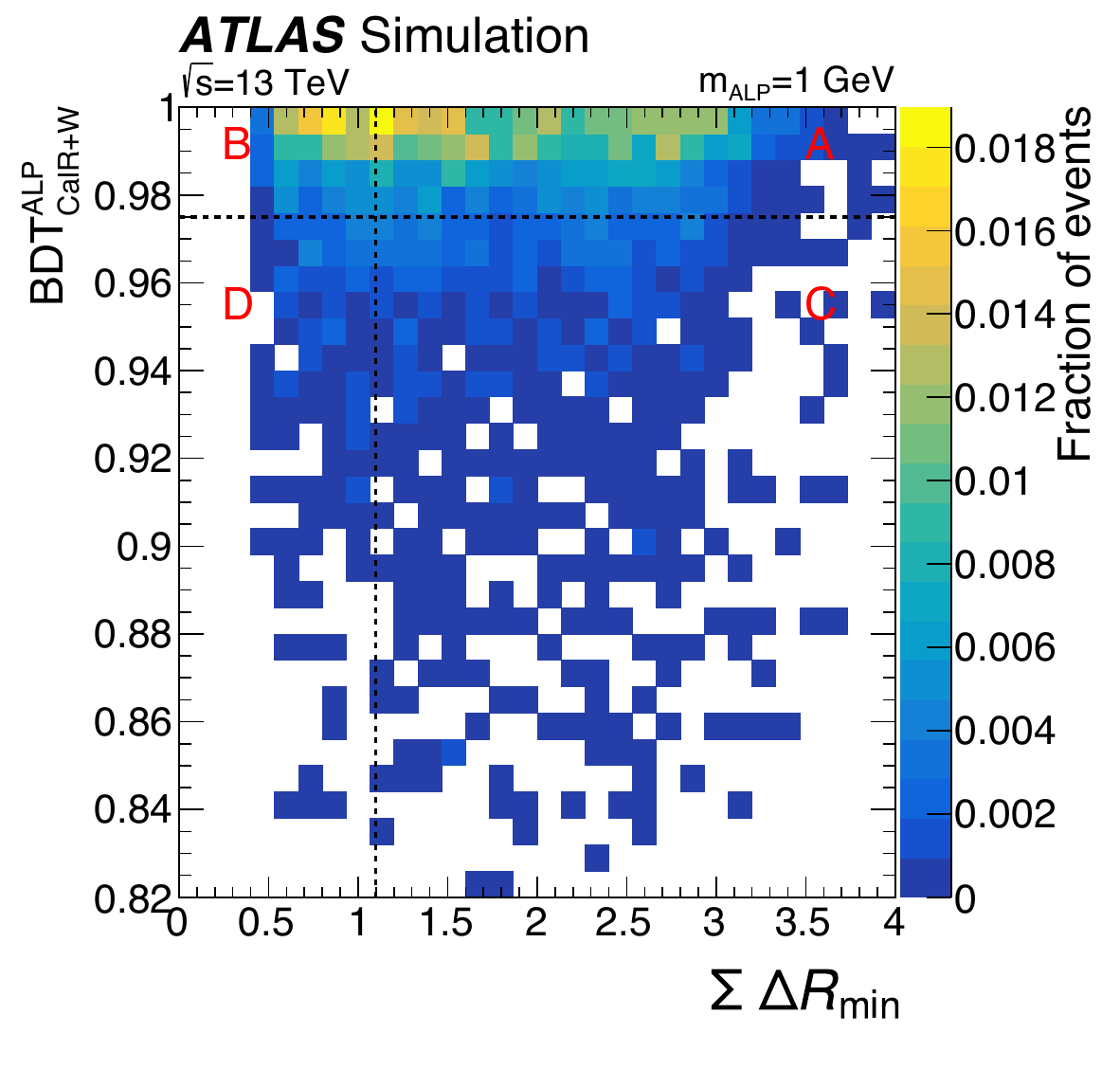}
} \subfloat[]{
\includegraphics[width=0.32\linewidth]{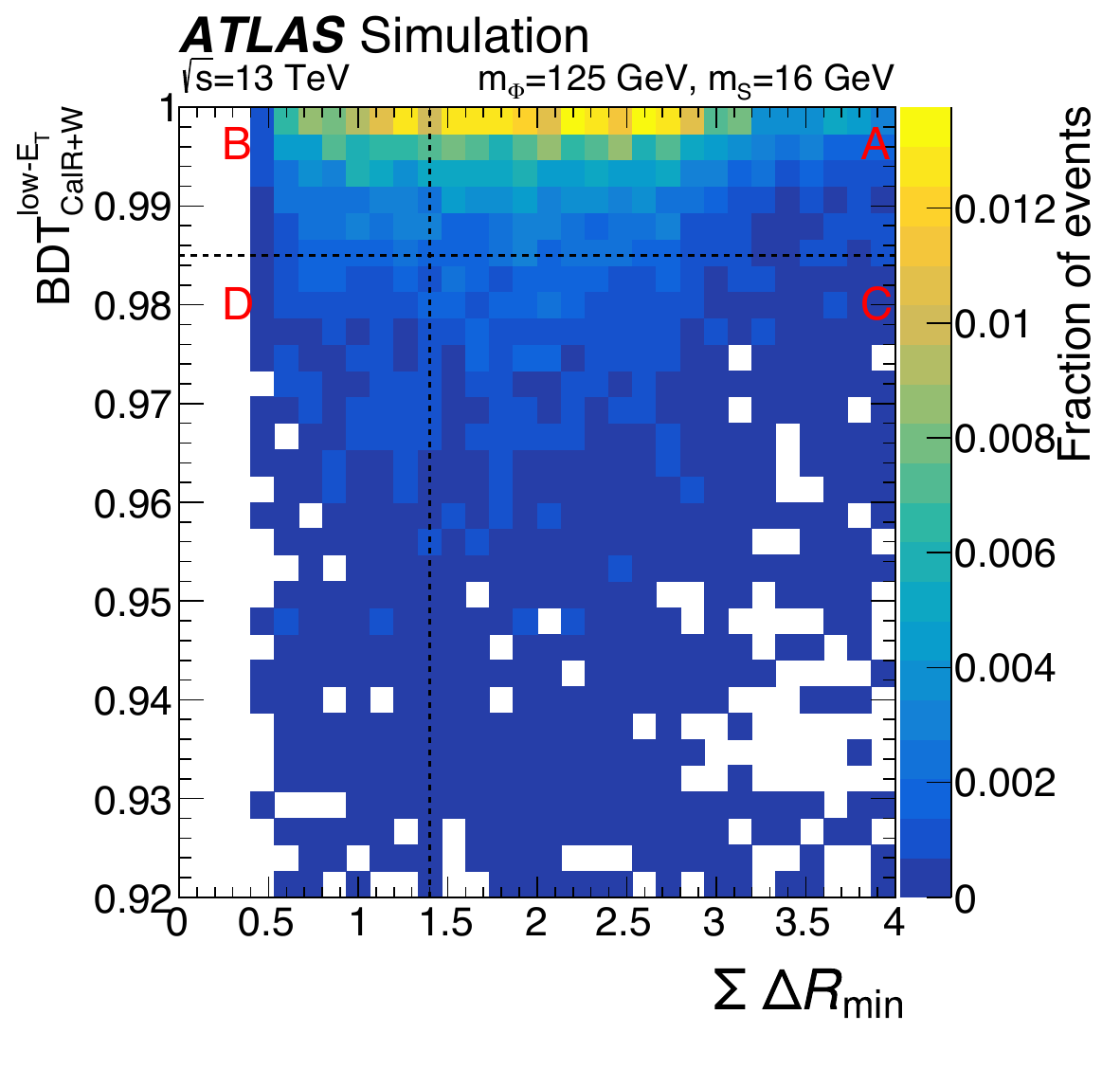}
}  \subfloat[]{
\includegraphics[width=0.32\linewidth]{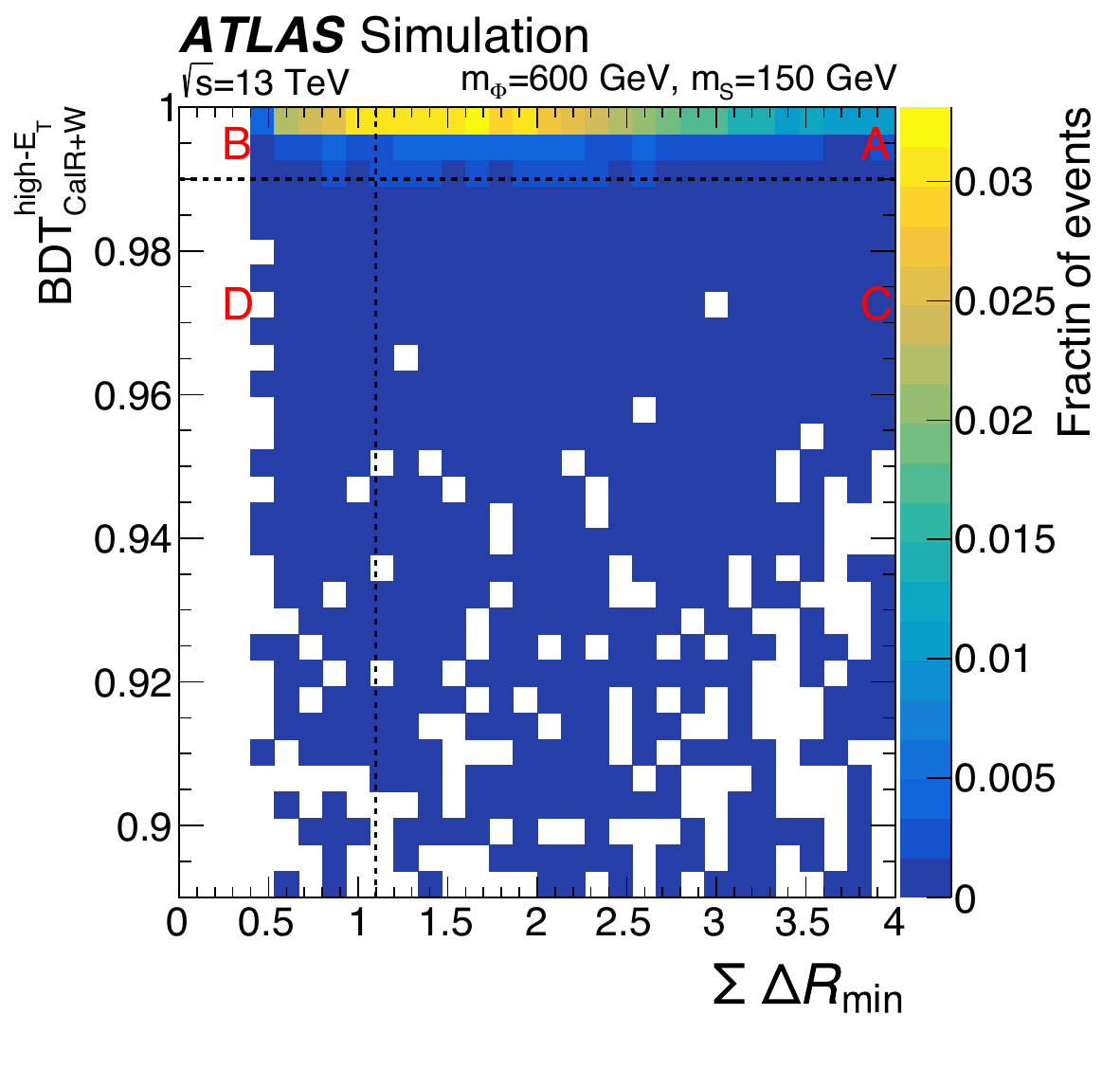}
}
\caption{Distribution of events in the two-dimensional plane of $\sum\Delta R_\textrm{min}$ vs. BDT score in the CalR+$W$ channel for (a--c) data and (d--f) example benchmark signal samples.
Final regions A, B, C and D are marked in red.
}
\label{fig:ABCD_W_Signal}
\end{figure}

\begin{figure}[t]
\centering
\subfloat[]{
\includegraphics[width=0.32\linewidth]{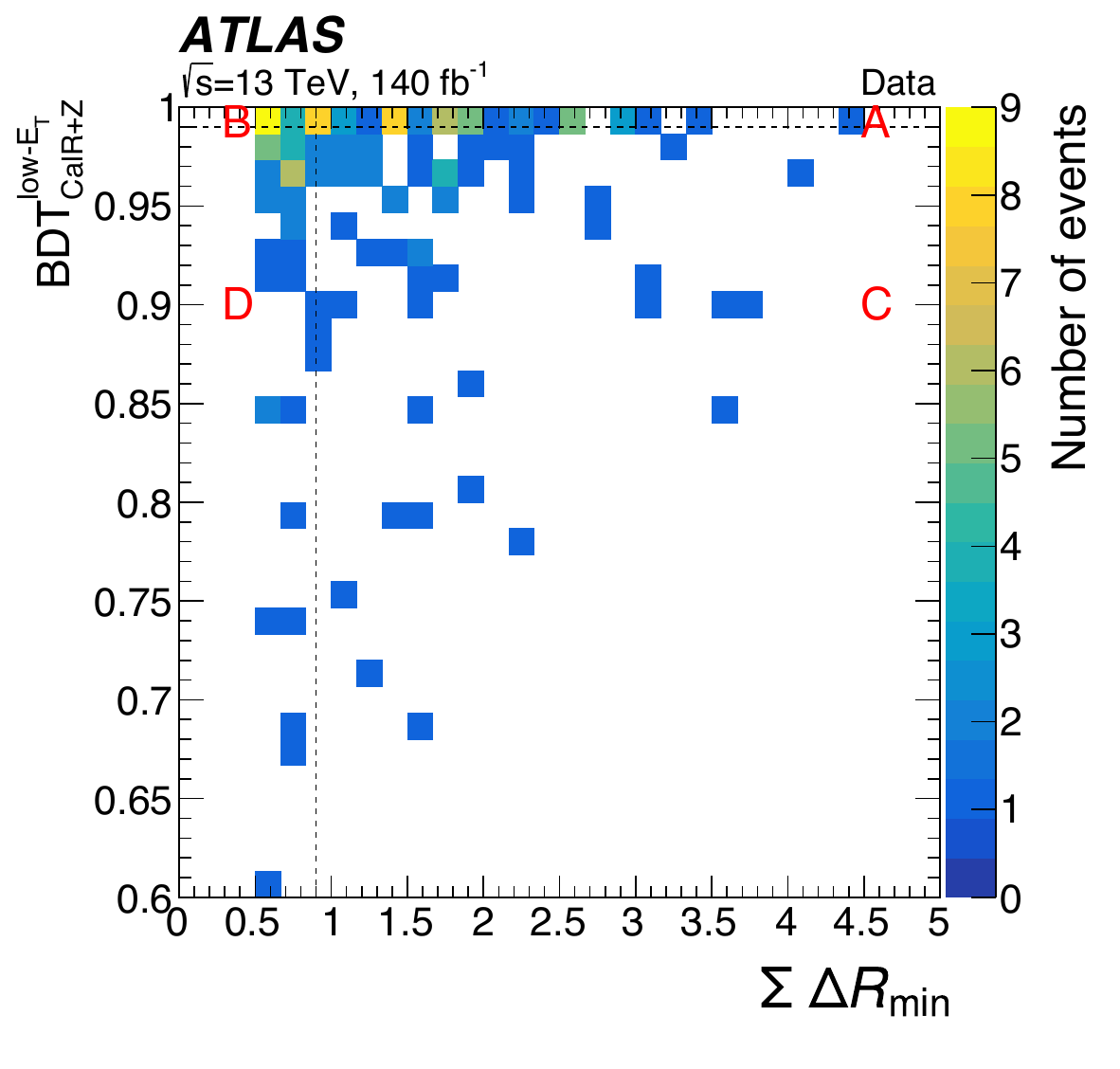}
} \subfloat[]{
\includegraphics[width=0.32\linewidth]{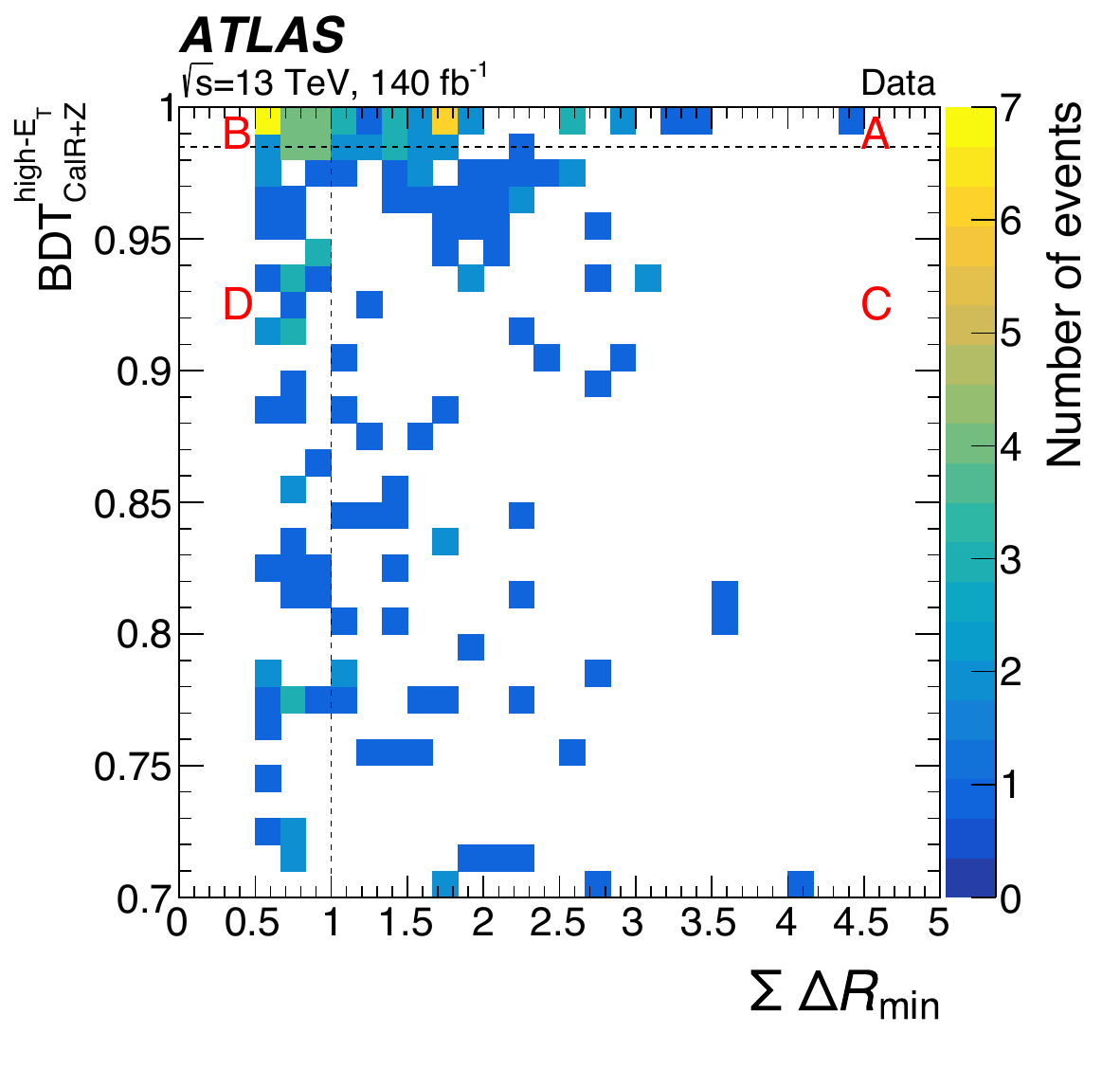}
}\\
\subfloat[]{
\includegraphics[width=0.32\linewidth]{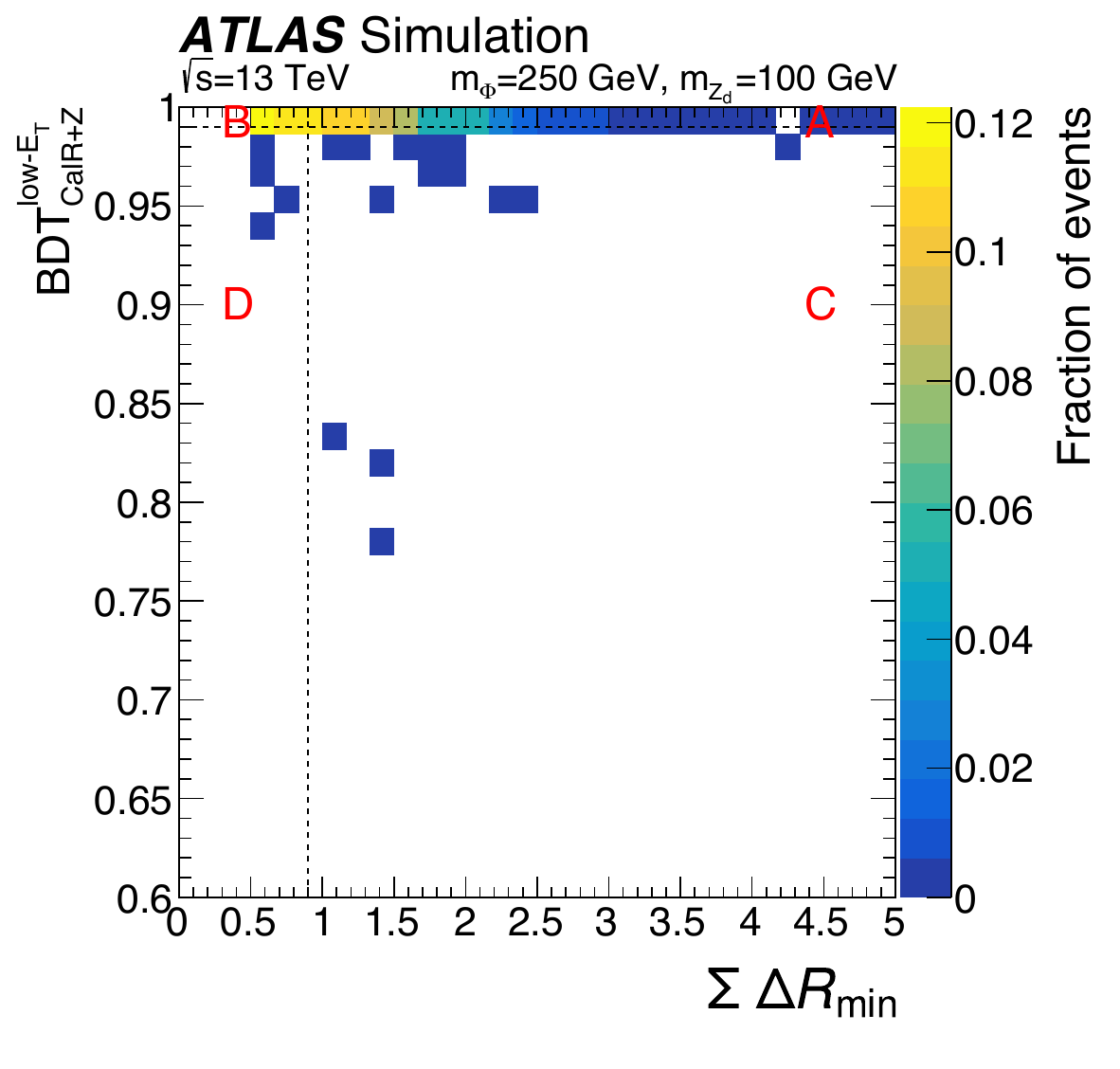}
}
\subfloat[]{
\includegraphics[width=0.32\linewidth]{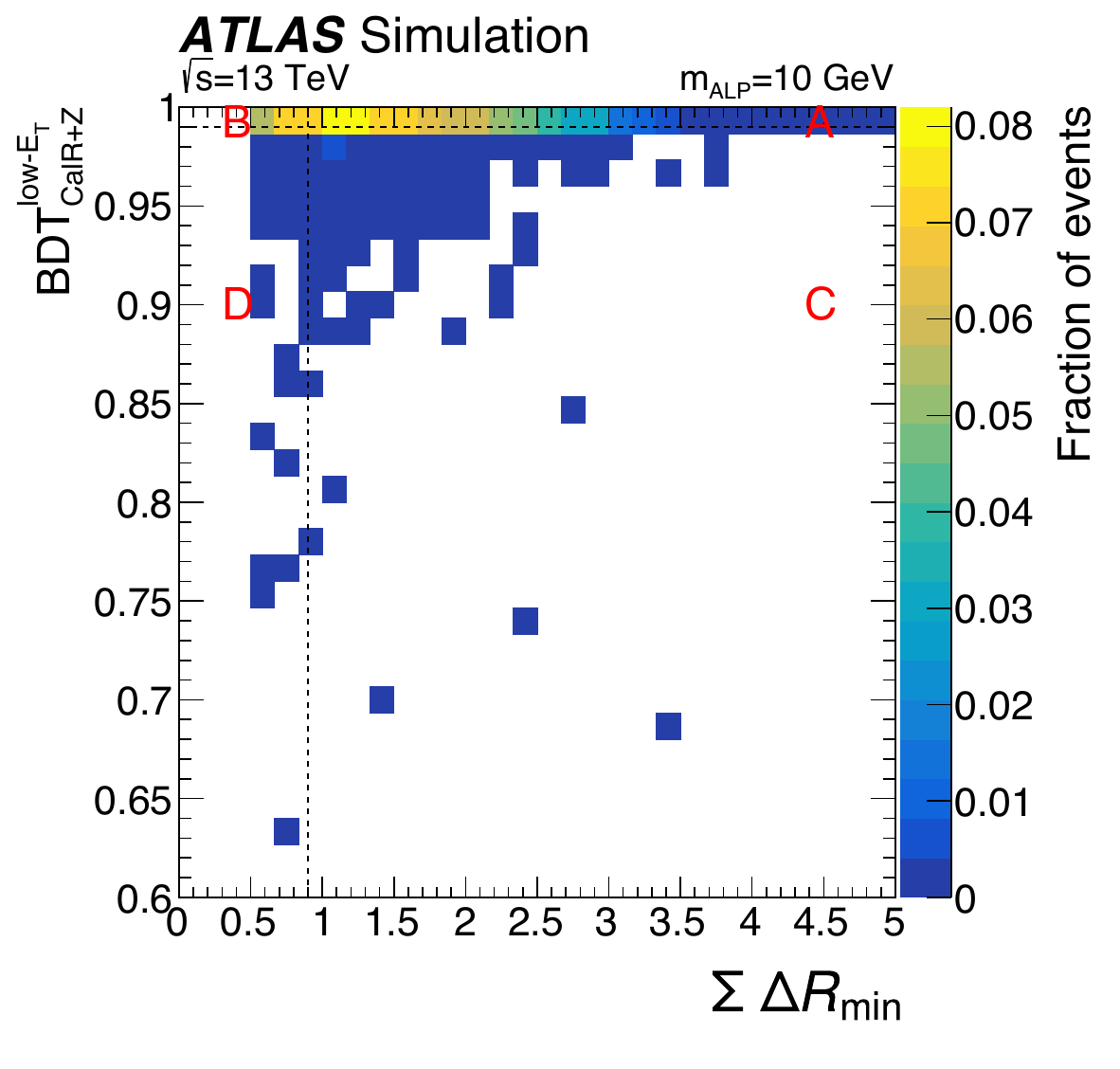}
} \subfloat[]{
\includegraphics[width=0.32\linewidth]{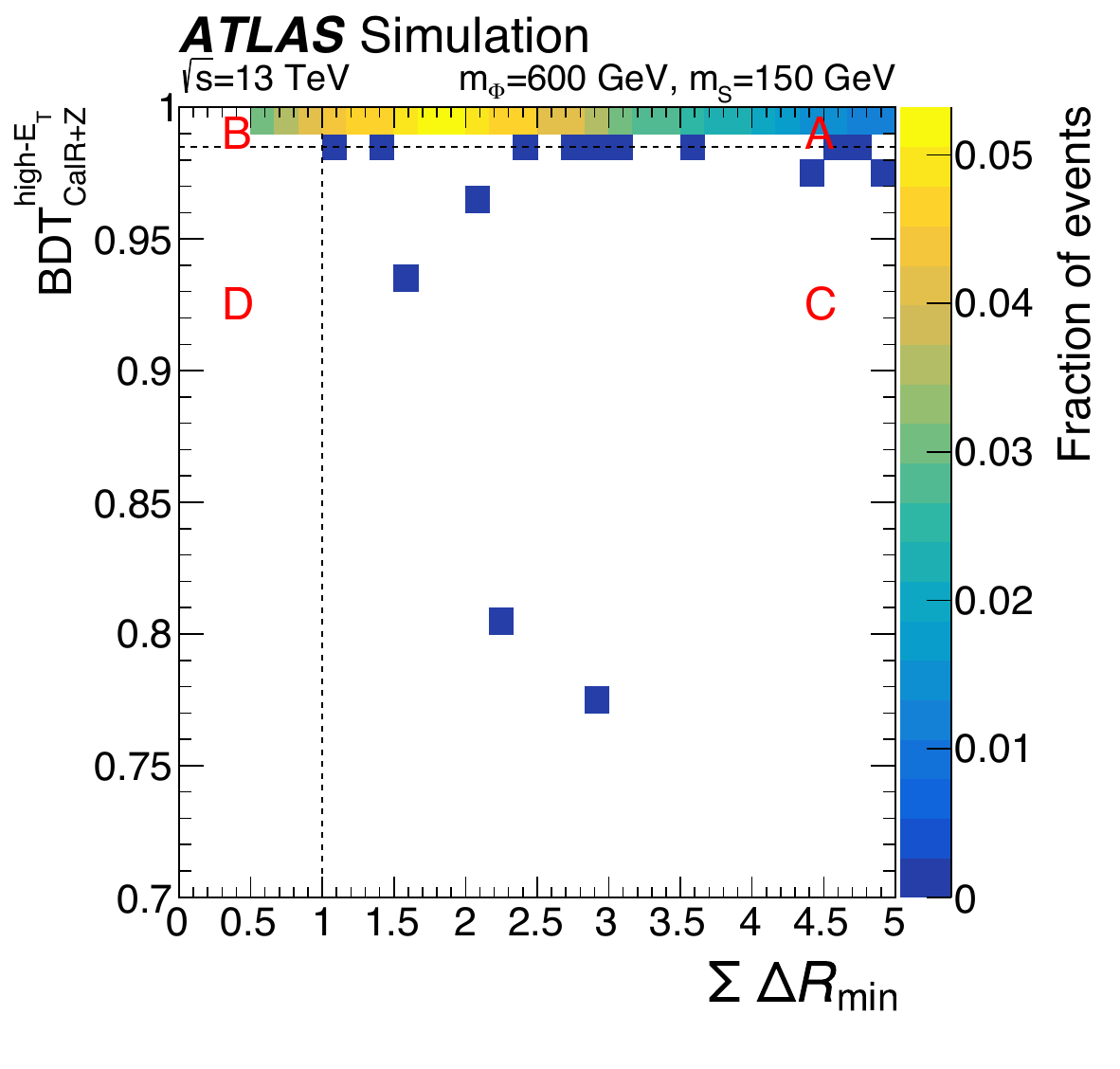}
}
\caption{Distribution of events in the two-dimensional plane of $\sum\Delta R_\textrm{min}$ vs. BDT score in the CalR+$Z$ channel for data (a, b) and (c--e) example benchmark signal samples. %
Final regions A, B, C and D are marked in red.
}
\label{fig:ABCD_Z_Signal}
\end{figure}

Figures~\ref{fig:ABCD_W_Signal} and~\ref{fig:ABCD_Z_Signal} show the ABCD planes for the CalR+W and CalR+Z channels respectively, for data and for one representative signal sample in each of the analysis selections. %

\section{Systematic uncertainties}
\label{sec:systematics}

The uncertainty in the integrated luminosity collected by ATLAS during LHC Run 2 is 0.83\%~\cite{DAPR-2021-01}, obtained using the LUCID-2 detector~\cite{LUCID2} for the primary luminosity measurement, complemented by measurements using the inner detector and calorimeters.
This affects the result during the conversion of the upper limit on the number of allowed signal events given the observed data, into an upper limit on the signal cross-section times branching fraction.

The background estimate is not affected by the typical modelling or experimental uncertainties, as it is data-driven for all three channels.
Validation tests for this method were performed in alternative ABCD planes, which were discussed in Section~\ref{sec:bkgs}.
The background estimate uncertainty is found to be negligible for all channels except for CalR+2J, in which it is estimated to be 5\%.

All remaining sources of uncertainty affect the signal efficiency estimate. The rest of this section exclusively discusses uncertainties of this type.
First, an uncertainty is assigned on the reweighting of events in simulation to match the observed distribution of pile-up in data.
The uncertainty in the pile-up reweighting of the reconstructed events in the MC simulation is estimated by comparing the distribution of the number of primary vertices in the MC simulation with the distribution in data as a function of the instantaneous luminosity. Differences between these distributions are reduced by scaling the mean number of interactions per bunch crossing in the MC simulation and $\pm1\sigma$ uncertainties are assigned to these scaling factors~\cite{DAPR-2021-01,LUCID2}. The effect on the signal event yields varies between 0.3\% and 7\% depending on the signal model and channel.

Next, the signal efficiency is affected by the jet energy scale (JES) and jet energy resolution (JER). These uncertainties are calculated using the prescription detailed in Ref.~\cite{Aaboud:2017jcu}, and are of the order of 0.1--15\% depending on the model. Since the displaced jets used in this analysis are non-standard due to their unusual electromagnetic energy fraction (EMF), the jet energy uncertainties were re-derived as a function of the EMF and $\eta$.
These additional uncertainties are found to have an effect of up to 4\% on the signals efficiencies. They are conservatively taken in quadrature with the standard jet energy uncertainties.

Since LLPs are used for triggering in the CalR+2J channel, an uncertainty is assigned on the signal trigger efficiency by studying the modelling of the key variables in the trigger: the jet energy, the CalRatio and the \pt of the jet's tracks. The level of agreement between data and the MC simulation for these variables in HLT- and offline-reconstructed quantities is evaluated. A tag-and-probe technique is applied to a pure sample of multijet events obtained using standard jet triggers in both the data and MC simulation. Scale factors that represent the degree of mismodelling in each variable are derived and then applied in an emulation of the CalRatio triggers. The change in yield relative to the nominal (unscaled) trigger emulation after the full analysis selection is taken as the size of the systematic uncertainty, which has values between 1\% and 6\% depending on the signal model.
In the CalR+$W$ and CalR+$Z$ channels, the triggering is performed using prompt leptons, and the offline selections on the lepton \pt are well above the trigger plateau. The lepton systematic uncertainty's effect on the signal event yields ranges from approximately 0.5\% to 0.9\%, depending on the specific benchmark model. This is small compared with other sources systematic error.

A systematic uncertainty in the machine-learning algorithms used in the analysis is included. Specifically, this uncertainty  accounts for potential mismodelling of input variables used in the per-jet NN tagger. It follows the same strategy as in Ref.~\cite{EXOT-2019-23} and is briefly summarised here.
In the dijet control regions, the distributions of the inputs of the per-jet NN are studied. The subsequent downstream ML algorithms are also explored in the relevant control regions. In most cases, the agreement is found to be good. Some variables exhibit modelling issues, as shown in the Auxiliary Material of Ref.~\cite{EXOT-2019-23}. Examples include longitudinal positions of track impact parameters, and hit and energy deposit timing information. These are caused by the fact that the spread of the beam-crossing time, as well as any instrumental cross-talk and timing offset calibrations are not simulated. The effect of such mismodelling is reduced by the adversarial training of the per-jet NN, which prevents it from exploiting differences between multijet events in simulation and in data. Nevertheless, the following procedure is used to gauge the residual effect on signal efficiency.
The mismodelling of problematic variables is encapsulated by a set of transfer factors. In an ensemble of pseudo-experiments, the per-jet NN and downstream ML variables are modulated by random  variations sampled from a Gaussian distribution with a mean of zero, and a width determined by the transfer factors determined above. The  signal efficiencies in each pseudo-experiment are recalculated with the modulated versions of the NN and ML algorithms.
The spread in the signal efficiencies across the ensemble of pseudo-experiments gives the uncertainty associated with this effect, which is estimated to be of the order of 5\% or smaller depending on the selection. An alternative method to evaluating the size of this uncertainty was studied. In this alternative method, control regions were defined by inverting the selections that define each signal region. The overall difference agreement of the per-event ML score distributions between data and simulation gave consistent values of the order of 5\%.

Small additional uncertainties arise from the electron and muon identification, calibration and reconstruction. The total systematic uncertainty associated with the leptons is below 1\% for all models, which is very small compared with all other sources of uncertainty. In the CalR+$W$, a small uncertainty of up to 1\% results from the missing transverse energy calibration.

Finally, an uncertainty due to the NLO-reweighting of the signal samples for the CalR+2J channel is obtained by comparing the NLO {\sc MadGraph} predictions for a 125 \GeV mediator mass with predictions at NNLO accuracy in QCD from {\sc Powheg Box} v2. This results in an additional signal efficiency uncertainty of 1\% to 6\% for most samples.
The CalR+$Z$/$W$ channels are not affected by this uncertainty.


\section{Results and statistical interpretation}
\label{sec:results}

The data-driven background estimation procedure described in Section ~\ref{sec:bkgs} is utilised in a simultaneous fit to assess if the data are compatible with the prediction or potential signal events. This approach follows a similar methodology as detailed in Ref.~\cite{EXOT-2019-23}, where a successful application of a comparable analysis method was previously demonstrated. An overall profile likelihood function is constructed from the product of the Poisson probabilities of
observing the number of events $N^{\textrm{obs}}_{X}$, given an expectation $N^{\textrm{exp}}_{X}$ in each region $X$, where $X$ = A, B, C or D.
The value of  $N^{\textrm{exp}}_{X}$ in each region is the sum of: the expected signal yield $N^{\textrm{sig}}_{X}$, given by the number of
simulated signal events entering region $X$ multiplied by the signal strength $\mu$ (the parameter of interest); and the expected background yield $N^{\textrm{bkg}}_{X}$. In the fit, the expected background yields are constrained to
obey the ABCD relation $N^{\textrm{bkg}}_\textrm{A} = (N^{\textrm{bkg}}_\textrm{B} \times N^{\textrm{bkg}}_\textrm{C})/N^{\textrm{bkg}}_\textrm{D}$. Since
the Poisson constraints only apply to $N^{\textrm{obs}}_{X}$ relative to $N^{\textrm{exp}}_{X}$, it follows that the background prediction may
change dynamically in the fit as a function of the signal strength $\mu$.
Additional constraint terms in the likelihood are included for the total signal uncertainty nuisance parameter.
The likelihood was implemented using the \texttt{pyhf} framework~\cite{pyhf_joss_local, pyhf}.

In the background-only hypothesis, the expected yields can only be adjusted for background components. Initially, without knowing the observed yield in region A, the expected background yields in regions B, C, and D match their observed counterparts, while region A reflects the naive ABCD relation. This initial estimate, known as the \emph{a priori} background estimate, assumes the observed value in region A is unknown, such as before unblinding the data. Once the observed yield in region A is included in the fit, the expected background yields in the other regions may shift away from their observed yields (within statistical uncertainties). This adjustment ensures that the ABCD relation holds for the \emph{expected} yields, even if the \emph{observed} yields deviate due to statistical fluctuations. This is called the \emph{a posteriori} background estimate. It corresponds to the case where the observed yield in region A is known (after unblinding).
The introduction of a signal component will  dynamically modify the allowed background prediction: for example allowing a potential excess in region A to be explained by a non-zero signal strength. The observed counts and the \emph{a priori} and \emph{a posteriori} results for each channel are summarised in Table~\ref{table:ABCD_table} for each region.

\begin{table}[ht!]
\caption{Application of the modified ABCD method to the final high-\et and low-\et selections.
The \emph{a priori} estimate refers to the ``pre-unblinding'' case, where the data in region A are ignored by removing the Poisson constraint in that region and the signal strength is fixed to zero.
This matches the simple $N^\mathrm{bkg}_\mathrm{A}=(N^\mathrm{bkg}_\mathrm{B} \times N^\mathrm{bkg}_\mathrm{C})/N^\mathrm{bkg}_\mathrm{D}$ relation.
The \emph{a posteriori} estimate refers to the ``post-unblinding'' case, including the observed data in region A in the background-only global fit. As can be seen, the predicted background in region A can increase substantially from the \emph{a priori} case, but only insofar as the adjusted yields in the other regions remain consistent with the observed yields within their statistical uncertainties.
Only statistical uncertainties are included in the quoted error on the background.
}
\begin{center}
\setlength{\tabcolsep}{0.2pc}{\small
\def\arraystretch{1.2}
\begin{tabular*}{\textwidth}{@{\extracolsep{\fill}}|l|r|r|r|r|}

\hline
\multicolumn{5}{|c|}{\textbf{CalR+2J channel}   }    \\ \hline

&  A                  &  B               &  C             & D             \\
\hline
Observed data                                & 92                  &  18               &  25213           &  4774          \\
Estimated background   \emph{a priori}                             & $95 \pm 23 $      &  $18\pm 4.2 $  &  $ 25210 \pm 160 $~ &  $ 4774 \pm 69$   \\
Fitted background      \emph{a posteriori}                          & $93 \pm 10 $     &  $18\pm 4.2 $  & $ 25210 \pm 160 $~ & $ 4774 \pm 69$    \\
\hline \hline

\multicolumn{5}{|c|}{\textbf{CalR+$W$ channel}   }    \\
\hline
$W$ ALP  selection              &  A                  &  B               &  C             & D             \\
\hline
Observed data                               & 27 & 23 & 122 & 82 \\
Estimated background   \emph{a priori}                          & $34.2 \pm 8.5$ & $23.0 \pm 4.8$ & $122 \pm 11$ & $82.0 \pm 9.1$ \\
Fitted background      \emph{a posteriori}                    & $29.3 \pm 4.4$ & $20.7 \pm 4.5$ & $120 \pm 11$ & $84.3 \pm 9.2$ \\
\hline

low-$E_{\mathrm{T}}$ $W$HS selection              &  A                  &  B               &  C             & D             \\
\hline
Observed data              & 59 & 53 & 155 & 155 \\
Estimated background   \emph{a priori}     &     $53.0 \pm 9.4$ & $53.0 \pm 7.3$ & $155 \pm 12$ & $155 \pm 12$ \\
Fitted background      \emph{a posteriori}  & $56.8 \pm 6.0$ & $55.2 \pm 7.4$ & $157 \pm 13$ & $153 \pm 12$   \\
\hline

high-$E_{\mathrm{T}}$ $W$HS selection              &  A                  &  B               &  C             & D             \\
\hline
Observed data                           & 33 & 21 & 261 & 220 \\
Estimated background   \emph{a priori}                         & $24.9 \pm 5.8$ & $21.0 \pm 4.6$ & $261 \pm 16$ & $220 \pm 15$ \\
Fitted background      \emph{a posteriori}                         & $29.6 \pm 4.2$ & $24.3 \pm 4.9$ & $264 \pm 16$ & $217 \pm 15$ \\
\hline \hline

\multicolumn{5}{|c|}{\textbf{CalR+$Z$ channel}   }    \\
\hline
low-$E_{\mathrm{T}}$ $Z$HS  selection             &  A                  &  B               &  C             & D            \\
\hline
Observed data    & 36 & 12 & 64 & 43 \\
Estimated background   \emph{a priori}           & $17.9 \pm 6.1$ & $12.0 \pm 3.5$ & $64.0 \pm 8.0$ & $43.0 \pm 6.6$ \\
Fitted background      \emph{a posteriori}          & $31.0 \pm 4.8$ & $17.0 \pm 3.9$ & $69.0 \pm 8.3$ & $38.0 \pm 6.2$ \\
\hline

high-$E_{\mathrm{T}}$ $Z$HS  selection              &  A                  &  B               &  C             & D             \\
\hline
Observed data          & 32 & 21 & 75 & 52 \\
Estimated background   \emph{a priori}       & $30.5 \pm 8.5$ & $21.0 \pm 4.6$ & $75 \pm 8.7$ & $52.0 \pm 7.2$ \\
Fitted background      \emph{a posteriori}    & $31.6 \pm 4.7$ & $21.3 \pm 4.6$ & $75.6 \pm 8.7$ & $51.4 \pm 7.2$ \\
\hline

\end{tabular*}
}
\end{center}
\label{table:ABCD_table}
\end{table}

No significant excess is observed in any of the signal regions. Upper limits on LLP production cross-section times branching fraction are extracted using the $CL_s$ method~\cite{Read:2002hq} with the ``alternative test statistic'' $\Tilde{q}$~\cite{Cowan:2010js} as implemented in \texttt{pyhf}.  The asymptotic approximation is used for all results. The results are obtained for the signal samples, which were generated with a given (arbitrary) LLP proper lifetime. The result can be extrapolated to other lifetimes by reweighting the LLP exponential decay distributions to each lifetime, and re-evaluating the signal efficiency. This procedure is identical to that described in Ref.~\cite{EXOT-2019-23}.

\begin{figure}[t!]
\centering
\subfloat[]{
\includegraphics[width=0.48\linewidth]{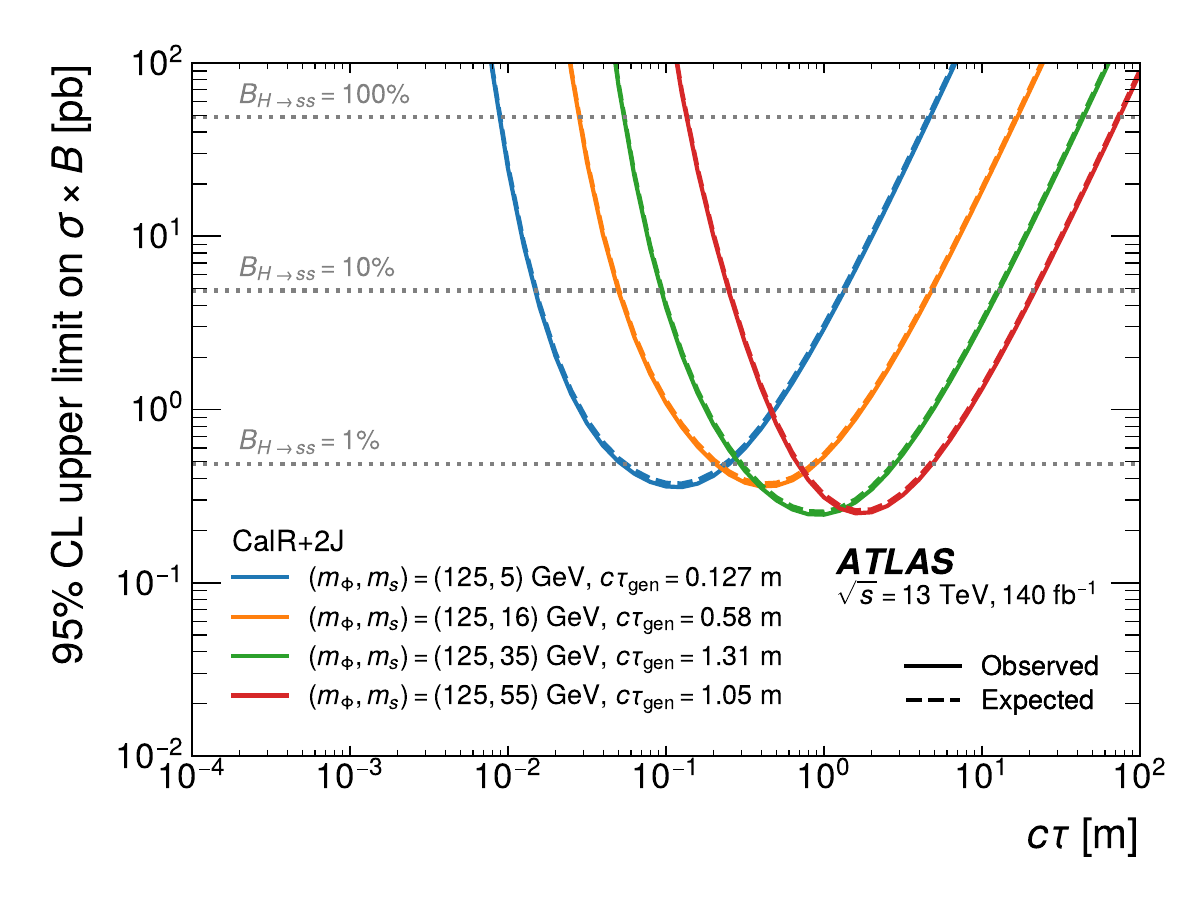}
}
\subfloat[]{
\includegraphics[width=0.48\linewidth]{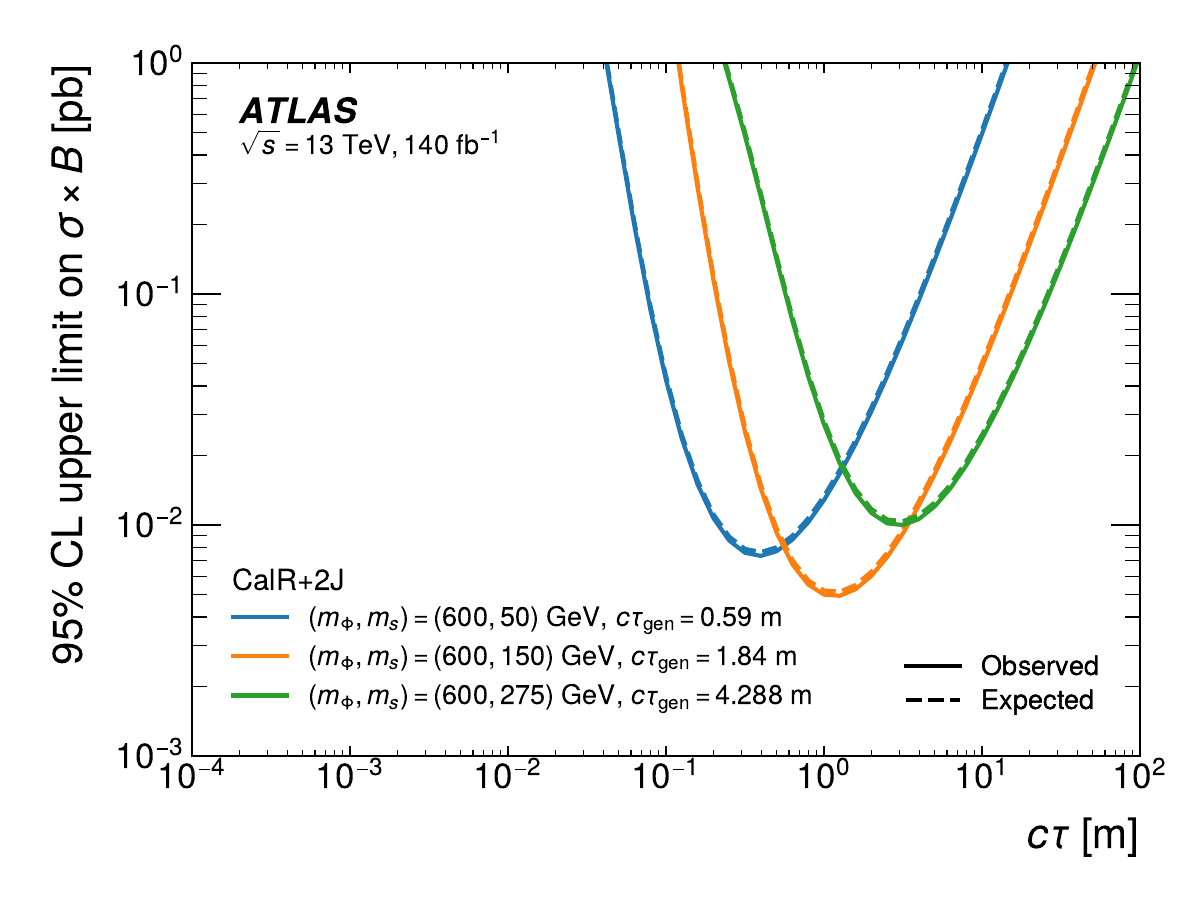}
}\\
\subfloat[]{
\includegraphics[width=0.48\linewidth]{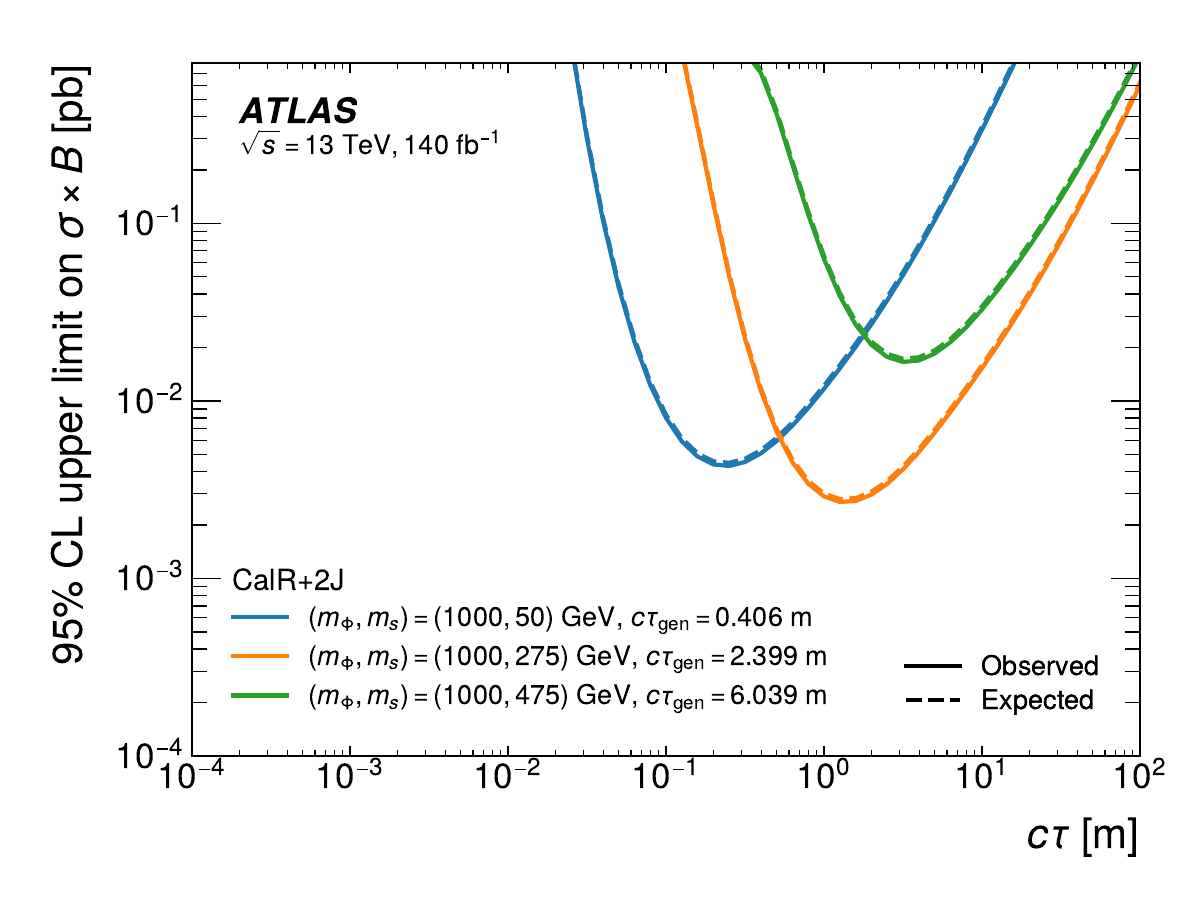}
}
\caption{Observed (solid line) and expected (dashed line) upper limits at the 95\% CL on the cross-section times branching fraction as a function of $c\tau$ for a selection of HS signal models in the CalR+2J channel for HS models with mediator masses of (a) 125 \GeV, (b) 600 \GeV and (c) 1000 \GeV. The $c\tau_{\textrm{gen}}$ values show where the curves were extrapolated from.}
\label{fig:cr2j_limits_sample}
\end{figure}

\begin{figure}[t!]
\centering
\subfloat[]{
\includegraphics[width=0.48\linewidth]{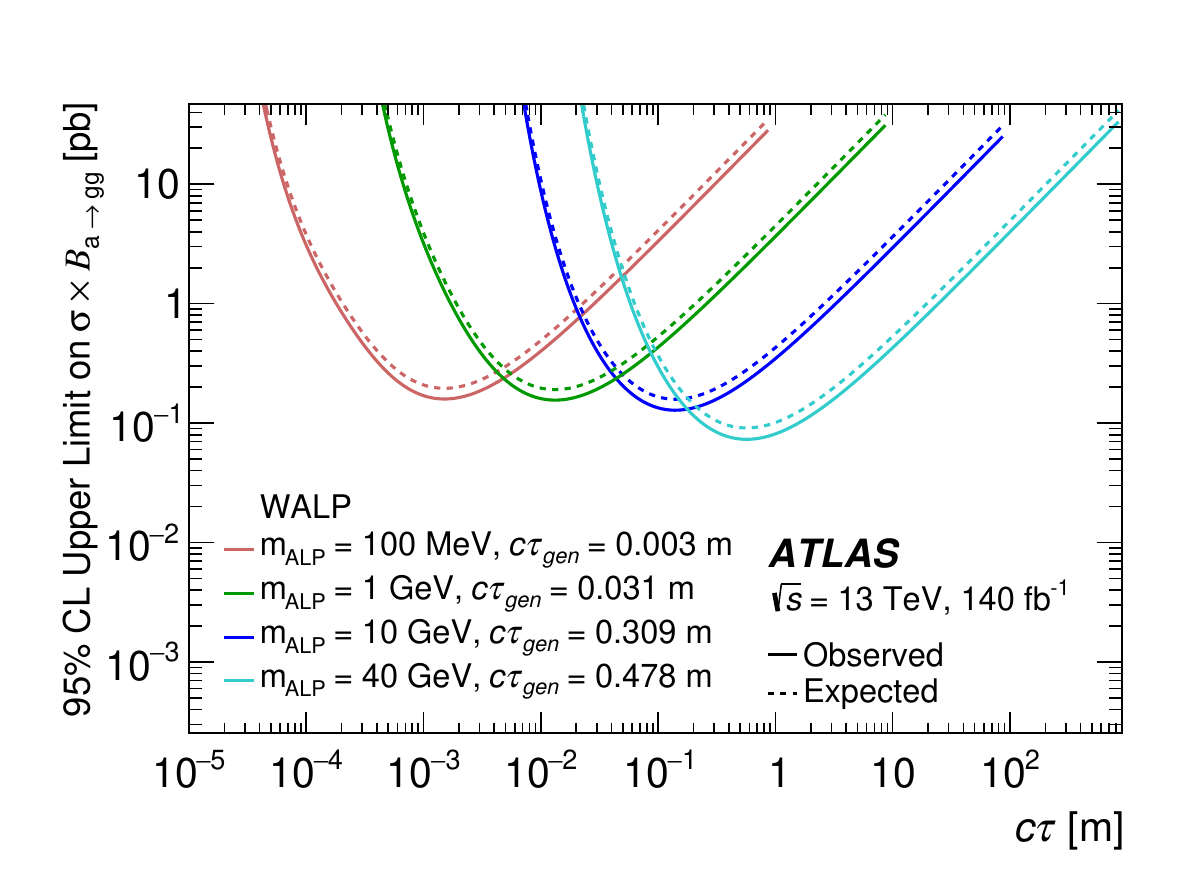}
}
\subfloat[]{
\includegraphics[width=0.48\linewidth]{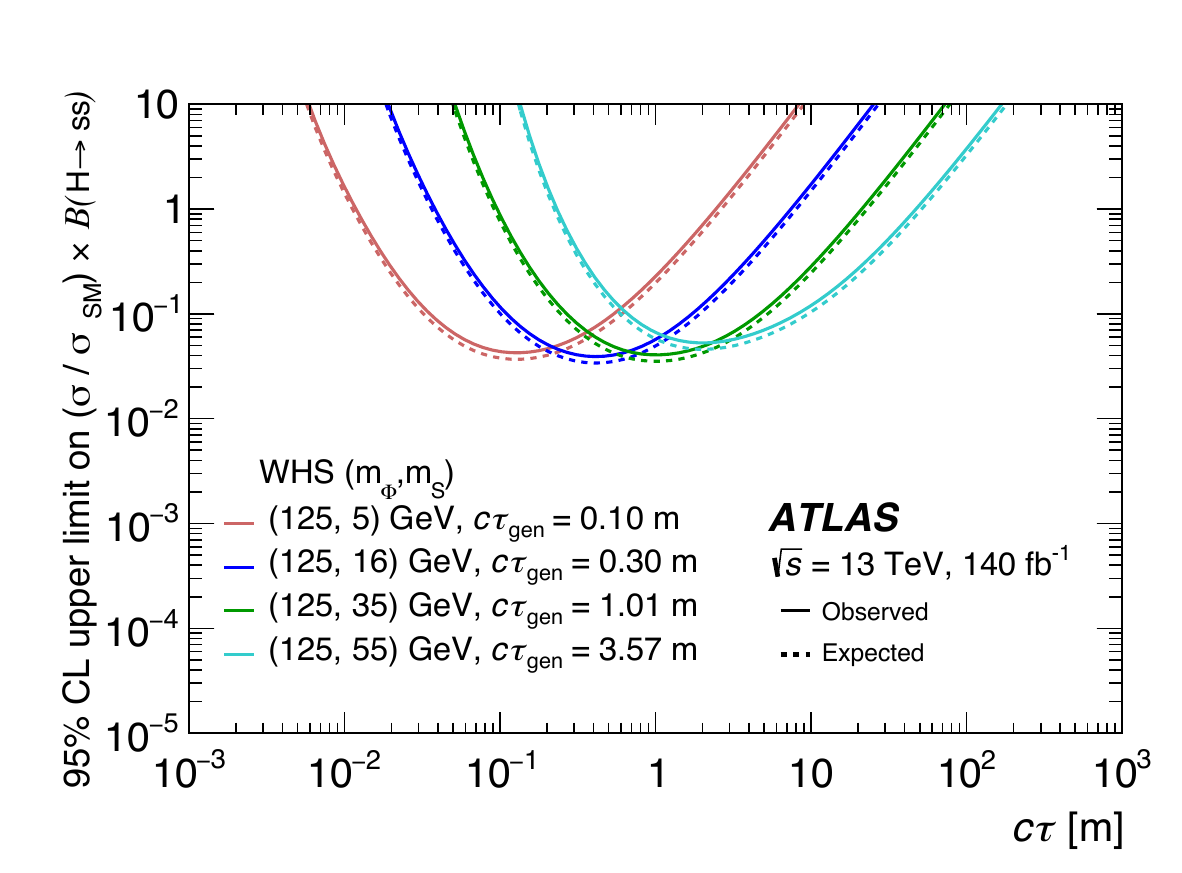}
}

\caption{Observed (solid line) and expected (dashed line) upper limits at the 95\% CL on the cross-section times branching fraction as a function of $c\tau$ for a selection of models targeted in the CalR+$W$ channel for (a) $W$ALP models and (b) $W$HS models with a SM Higgs boson mediator. The $c\tau_{\textrm{gen}}$ values show where the curves were extrapolated from.}
\label{fig:CRW_limits_sample}
\end{figure}

\begin{figure}[t!]
\centering
\subfloat[]{
\includegraphics[width=0.48\linewidth]{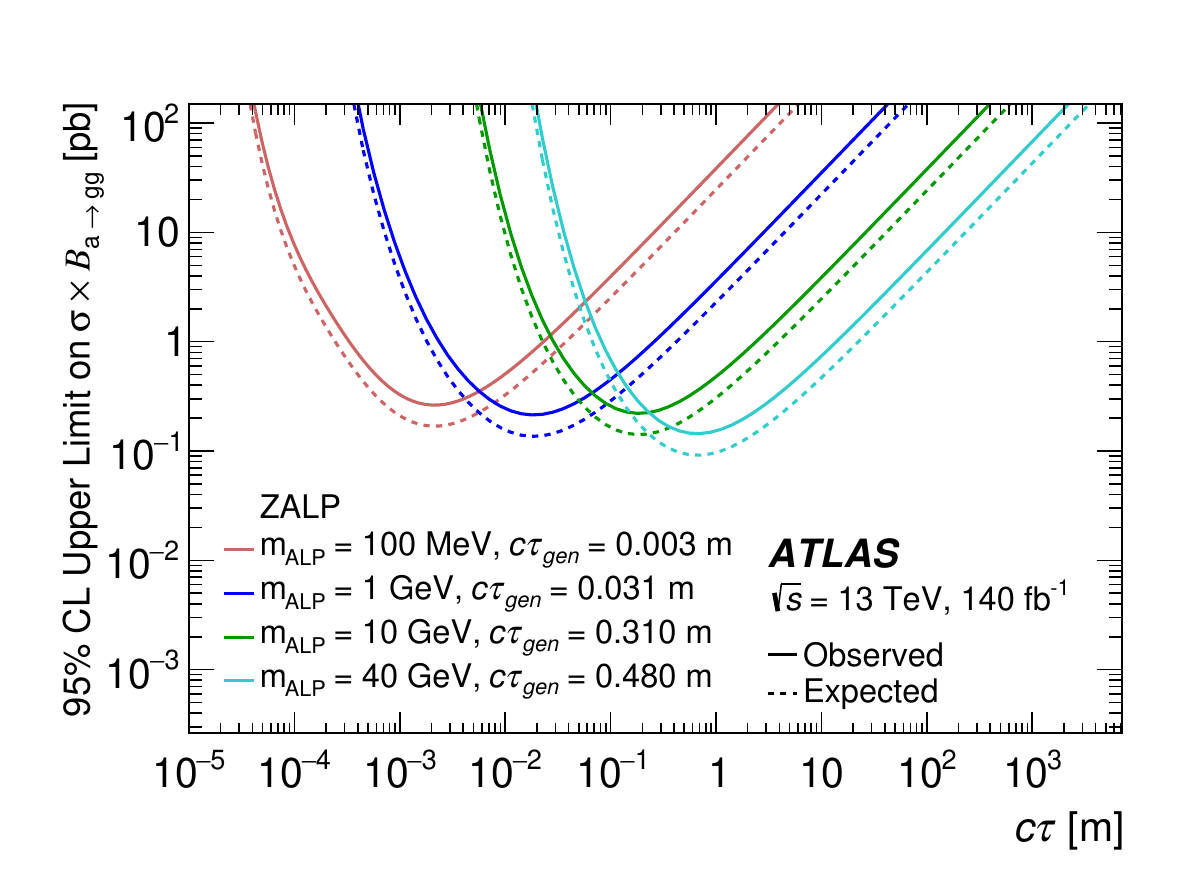}
}
\subfloat[]{
\includegraphics[width=0.48\linewidth]{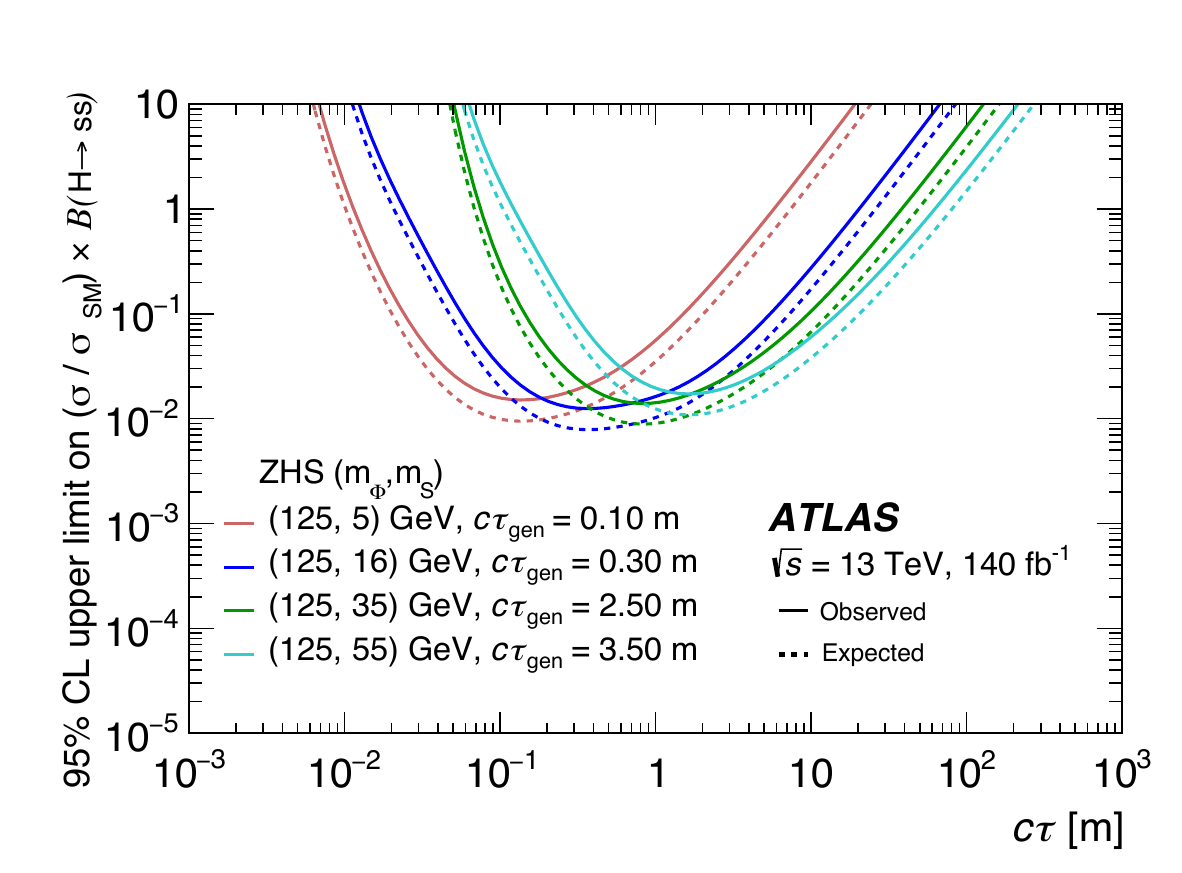}
}\\
\subfloat[]{
\includegraphics[width=0.48\linewidth]{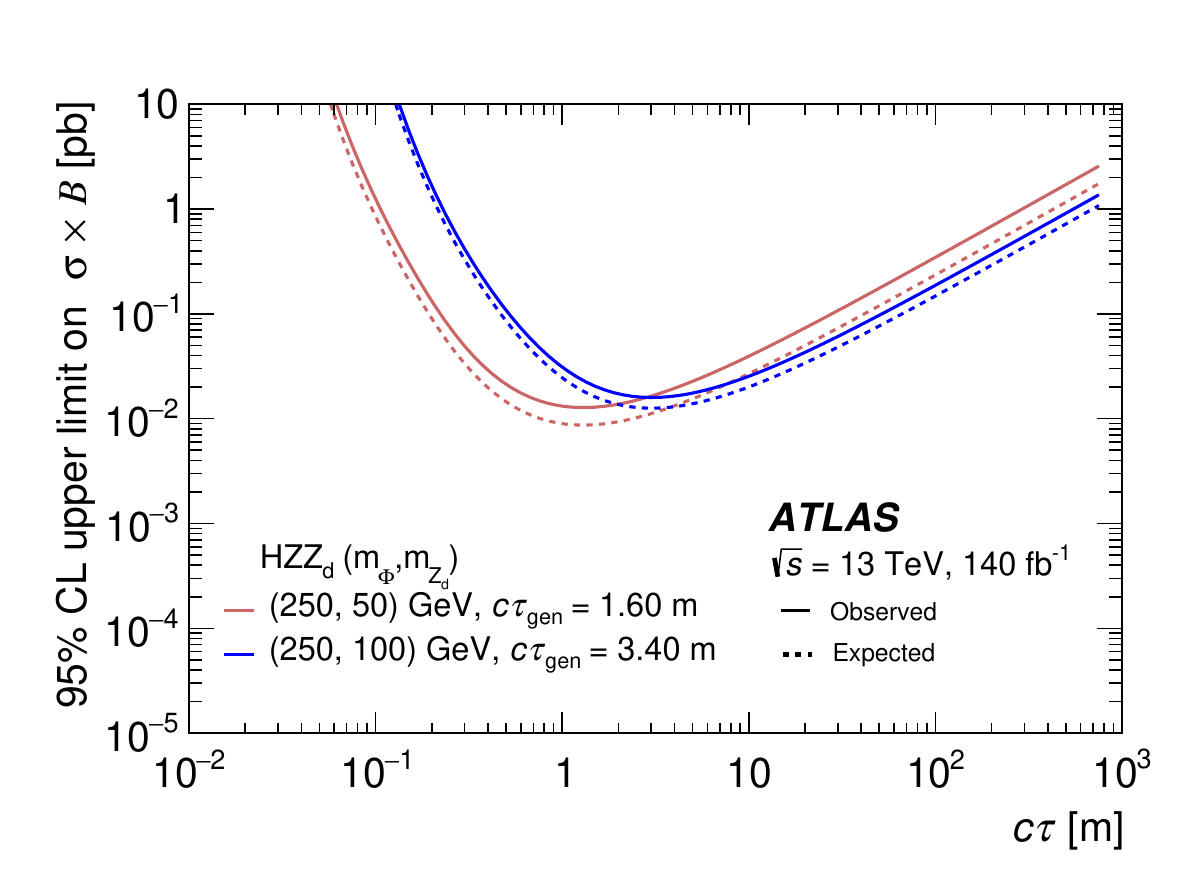}
}
\caption{Observed (solid line) and expected (dashed line) upper limits at the 95\% CL on the cross-section times branching fraction as a function of $c\tau$ for a variety of models probed by the CalR+$Z$ channel for (a) $Z$ALP models, (b) $Z$HS models with the SM Higgs boson as a mediator, and (c) $HZZ_d$ models. The $c\tau_{\textrm{gen}}$ values show where the curves were extrapolated from.}
\label{fig:CRZ_limits_sample}
\end{figure}

Examples of the resulting 95\% confidence level (CL) upper limits on the cross-section times branching fraction for HS models based on the CalR+2J channel are shown in Figure~\ref{fig:cr2j_limits_sample}.
Figure~\ref{fig:CRW_limits_sample} shows a selection of limits obtained using the CalR+$W$ channel, constraining the $W$ALP model and the HS model in the associated $W$ boson production mode.
The limits derived from the CalR+$Z$ channel are presented in Figure \ref{fig:CRZ_limits_sample} and constrain the $Z$ALP, $HZZ_{d}$ and $Z$HS models.
For the case of $m_{\Phi}=125$ \GeV, the limits are scaled to the relevant SM cross-section of Higgs boson production.

\begin{figure}[t!]
\centering
\subfloat[]{
\includegraphics[width=0.45\linewidth]{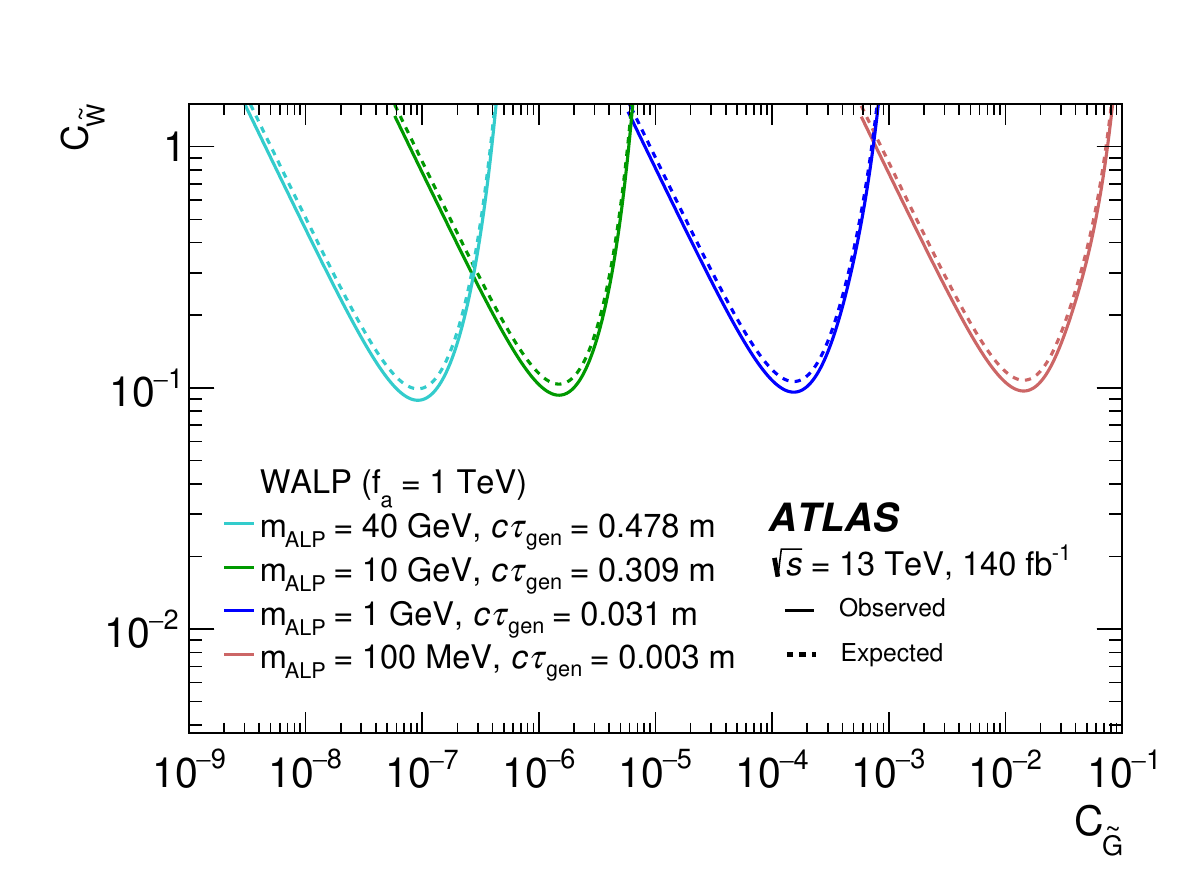}
}
\subfloat[]{
\includegraphics[width=0.45\linewidth]{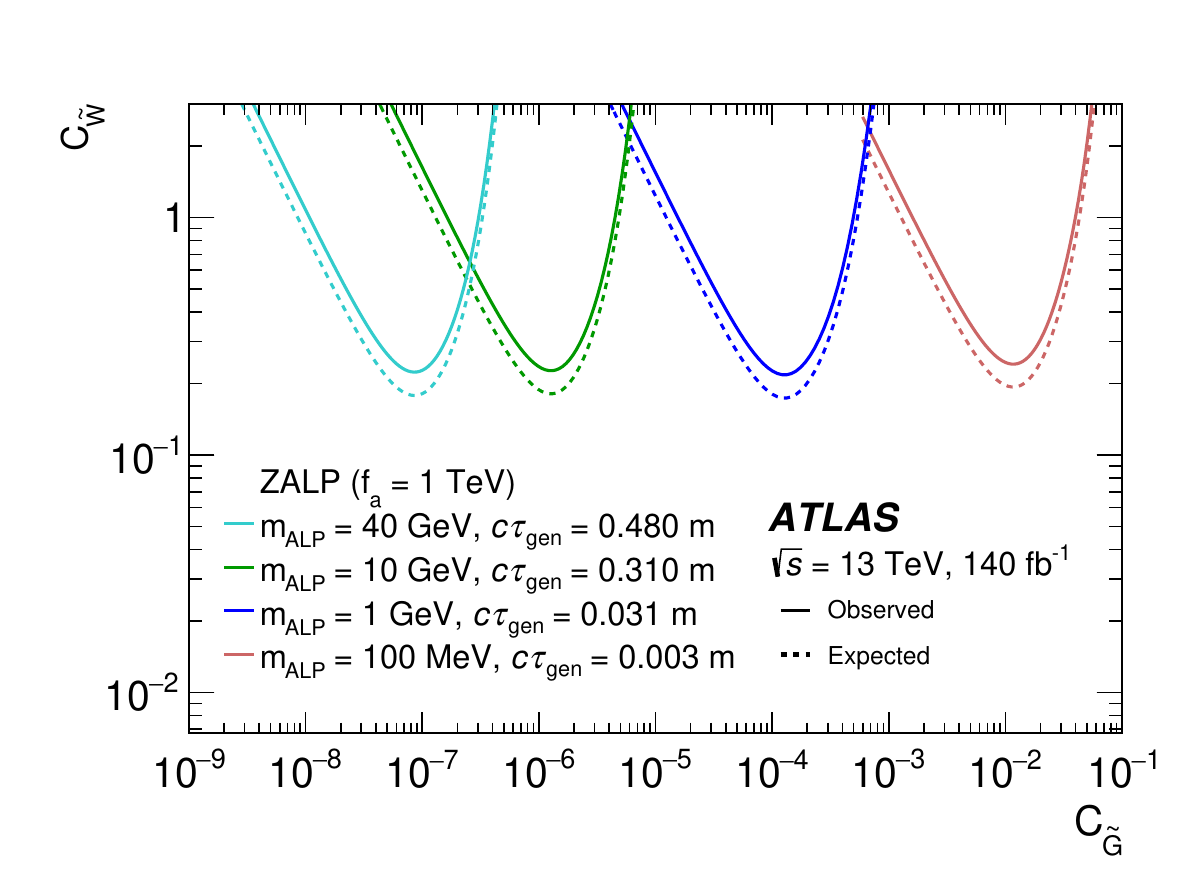}
}
\caption{Summary of the observed (solid line) and expected (dashed line) upper limits at the 95\% CL on $C_{\tilde{G}}$ as a function of $C_{\tilde{W}}$  for a variety of ALP mass hypotheses probed by the CalR+$R$ and Cal+$Z$ channels for (a) $W$ALP and (b) $Z$ALP models. The $c\tau_{\textrm{gen}}$ values show where the curves were extrapolated from.}
\label{fig:ZALP_summary_limits_CGCW}
\end{figure}

\begin{figure}[t!]
\centering
\subfloat[]{
\includegraphics[width=0.45\linewidth]{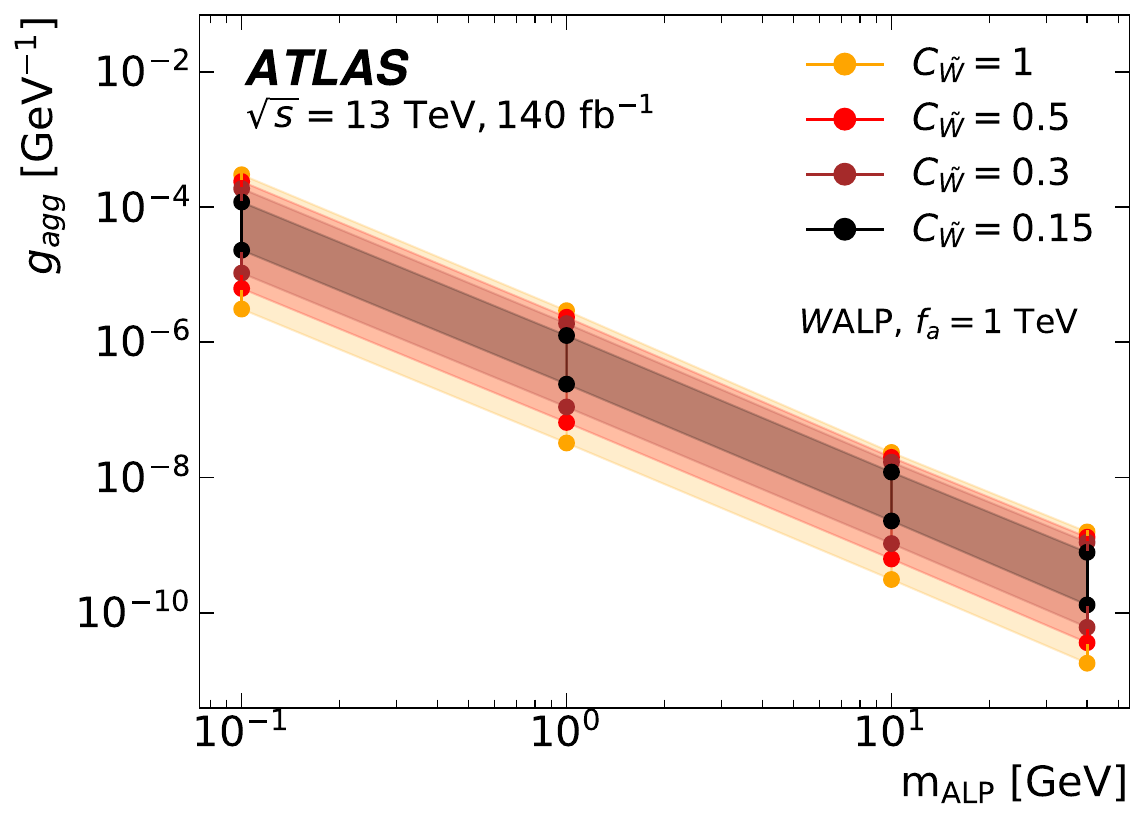}
}
\subfloat[]{
\includegraphics[width=0.45\linewidth]{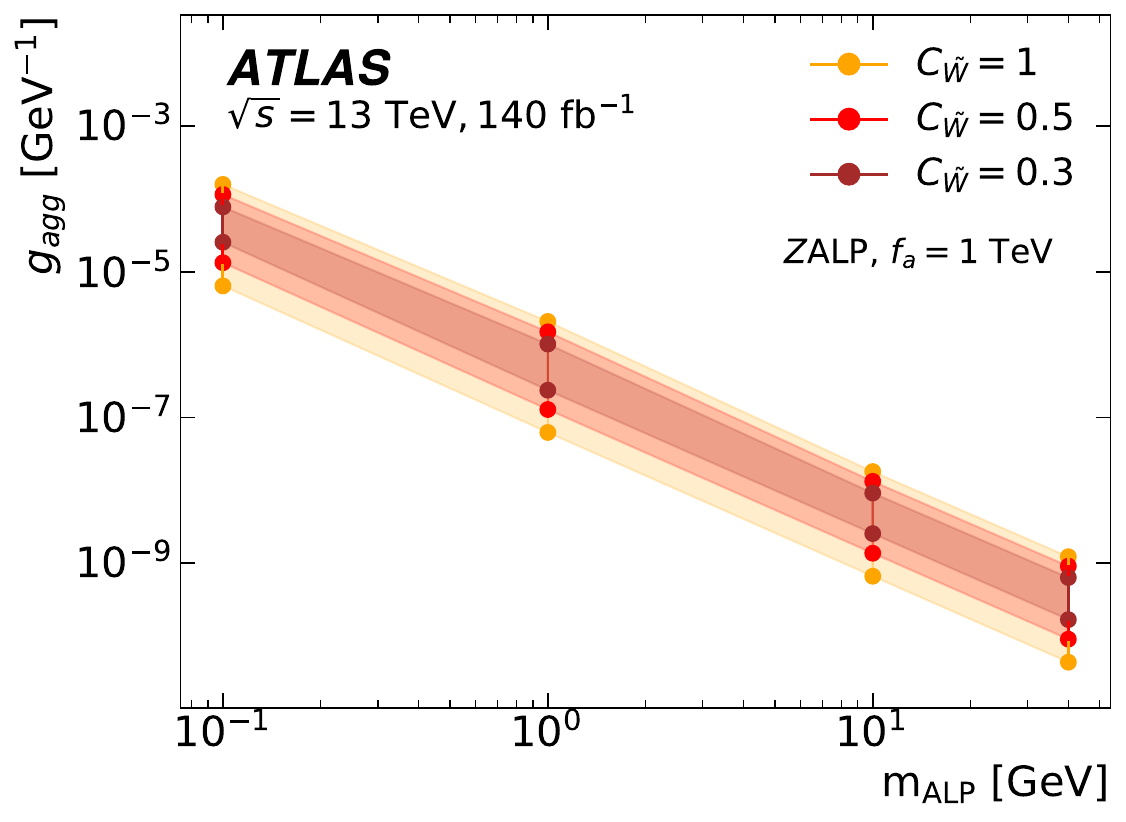}
}
\caption{
Regions in the ALP mass versus its coupling to gluons scaled by the ALP energy scale $g_{agg}=4C_{\tilde{G}}/f_a$ excluded at 95\% CL, for $C_{\tilde{W}}=1$, $C_{\tilde{W}}=0.5$, $C_{\tilde{W}}=0.3$ and $C_{\tilde{W}}=0.15$, in the (a) $W$ALP and (b) $Z$ALP models. The contours are obtained by interpolating between the exclusions for different LLP masses considered in the analysis, taking the logical union of the regions excluded for each mass.}
\label{fig:ZALP_summary_limits_CggMa}
\end{figure}

The constraints obtained by CalR+2J for HS models are strongest for proper lifetimes $c\tau$ of order of 1--2~m. For the particular case where the SM Higgs boson is the mediator, assuming a production cross-section for gluon--gluon fusion of 48~pb, the branching fraction of Higgs bosons to hadronically decaying scalar LLPs is constrained to below 1\% in the mean proper decay length range of 30~cm to 4.5~m, depending on the mass of the LLP\@. This represents an improvement of a factor of approximately three relative to the results in the previous search~\cite{EXOT-2019-23}. Specifically, this improvement comes from relaxing a requirement on the CalRatio of the two most signal-like jets (which was detrimental to the events featuring the resolved topology but necessary to control backgrounds), and instead exploiting additional jet information, allowing a substantial background reduction while maintaining signal efficiency for low mediator mass samples.

Still considering the HS model, the CalR+$Z$ and CalR+$W$ channels provide the first constraints on hadronically decaying LLPs in the calorimeter where the production of a scalar mediator is in association with a vector boson. Higgs boson branching fractions to LLPs above 50\% in this production mode are excluded for mean proper decay lengths in the~cm to~m range. While the derived constraints in terms of Higgs boson branching fractions are not as strong as those derived in the CalR+2J channel or in Ref.~\cite{EXOT-2019-23}, they provide a complementary constraint in a different production mode. A combination of the HS results between the three channels is not performed, since the gluon--gluon fusion cross-section is overwhelmingly large compared with the associated vector boson production cross-section and consequently any combined limit would be fully dominated by the  CalR+2J channel results.

Constraints are set by ATLAS on photophobic ALP models for the first time using the CalR+$W$ and CalR+$Z$ channels, with cross-sections above 0.1~pb excluded in the 0.1~mm--10~m range.
ALP exclusion limits in terms of cross-section and proper lifetime can be interpreted for the relevant model parameters. The width of the ALP decay into gluons is given by:
\begin{equation}
\Gamma_{agg}=\frac{2}{\pi f_a^{2}}C^2_{\tilde{G}}m_{\textrm{ALP}}^3,
\end{equation}
where $f_a$ is the energy scale of the ALP and $m_\textrm{ALP}$ its mass. Therefore, for a given mass and scale, the ALP proper lifetime $c\tau$ depends on the inverse square of the gluon coupling $c\tau \propto C^{-2}_{\tilde{G}}$. Additionally, the cross-section of the simulated process is determined by the coupling of the ALP with the SM weak-sector. For a photophobic ALP, this implies that $\sigma \propto C^2_{\tilde{W}}$. The resulting exclusion limits on $C_{\tilde{G}}$ and $C_{\tilde{W}}$ are presented in Figure~\ref{fig:ZALP_summary_limits_CGCW}. A summary of these results is presented in the plane of ALP mass versus its coupling to gluons in Figure~\ref{fig:ZALP_summary_limits_CggMa}, where the coupling to gluons is scaled by the ALP energy scale $g_{agg}=4C_{\tilde{G}}/f_a$.

Finally, new constraints on the $ZZ_d$ model are set, which are more sensitive than previous results~\cite{EXOT-2017-24} by an order of magnitude, with production cross-sections of a dark photon with a $Z$ boson above 0.1~pb excluded for $Z_d$ proper decay lengths in the 20~cm to 50~m range, depending on the mediator and $Z_d$ masses. The improvement comes from the fact that this work uses a dataset approximately four times larger than the one studied in Ref.~\cite{EXOT-2017-24}, and major improvements in displaced jet identification efficiency and background rejection thanks to the per-jet NN.


\FloatBarrier

\section{Conclusion}
\label{sec:conclusions}

This paper describes a search for hadronically decaying long-lived particles giving rise to displaced jets in the ATLAS hadronic calorimeter, using the ATLAS Run~2 dataset of 140 fb$^{-1}$ of \(pp\) collisions at $\sqrt{s}=13\TeV$. The analysis considers three topologies: pair-produced LLPs, one of which leads to a single displaced jet in the calorimeter, the other, having low boost, yielding two resolved jets; and single- or pair-produced LLPs in association with a $W$ or $Z$ vector boson, which decays leptonically.
For all three analysis channels a per-jet neural network is used to distinguish signal-like jets from jets from Standard Model processes or beam-induced backgrounds. Further per-event machine-learning discriminants are then used to define regions sensitive to a variety of LLP signals. In each case, a data-driven background estimate is used.
No significant excess of events is observed in any of the channels. Constraints on the production cross-section times branching fraction at 95\% confidence level are set for several benchmark models. The results for Hidden Sector models exclude branching fractions of Higgs boson mediators to neutral scalar LLPs above 1\% for LLP decay lengths between 30~cm and 4.5~m. These results extend those set by previous ATLAS results, improving the limits by a factor of approximately three, by exploiting additional jet information to more efficiently reject background events. Limits are also set on dark photons produced in association with a $Z$ boson, with production cross-sections above 0.1~pb excluded for proper decay lengths between 20~cm and 50~m. This improves upon previous ATLAS results by an order of magnitude, thanks to a larger dataset and the use of machine-learning discriminants. Finally, constraints are set on photophobic axion-like particle models produced in association with a $W$ or $Z$ boson for the first time, with cross-sections above 0.1~pb excluded in the 10~mm--10~m range.
This analysis is part of a wider programme of searches for LLPs with the ATLAS experiment, and has unique sensitivity to LLP laboratory decay lengths of the order of 1~m. It represents the first time that signatures involving neutral hadronically decaying LLPs with prompt objects are targeted in that decay range with the aid of machine-learning discriminants.


\section*{Acknowledgements}

%
%

%
%

We thank CERN for the very successful operation of the LHC and its injectors, as well as the support staff at
CERN and at our institutions worldwide without whom ATLAS could not be operated efficiently.

The crucial computing support from all WLCG partners is acknowledged gratefully, in particular from CERN, the ATLAS Tier-1 facilities at TRIUMF/SFU (Canada), NDGF (Denmark, Norway, Sweden), CC-IN2P3 (France), KIT/GridKA (Germany), INFN-CNAF (Italy), NL-T1 (Netherlands), PIC (Spain), RAL (UK) and BNL (USA), the Tier-2 facilities worldwide and large non-WLCG resource providers. Major contributors of computing resources are listed in Ref.~\cite{ATL-SOFT-PUB-2023-001}.

We gratefully acknowledge the support of ANPCyT, Argentina; YerPhI, Armenia; ARC, Australia; BMWFW and FWF, Austria; ANAS, Azerbaijan; CNPq and FAPESP, Brazil; NSERC, NRC and CFI, Canada; CERN; ANID, Chile; CAS, MOST and NSFC, China; Minciencias, Colombia; MEYS CR, Czech Republic; DNRF and DNSRC, Denmark; IN2P3-CNRS and CEA-DRF/IRFU, France; SRNSFG, Georgia; BMBF, HGF and MPG, Germany; GSRI, Greece; RGC and Hong Kong SAR, China; ISF and Benoziyo Center, Israel; INFN, Italy; MEXT and JSPS, Japan; CNRST, Morocco; NWO, Netherlands; RCN, Norway; MNiSW, Poland; FCT, Portugal; MNE/IFA, Romania; MSTDI, Serbia; MSSR, Slovakia; ARIS and MVZI, Slovenia; DSI/NRF, South Africa; MICIU/AEI, Spain; SRC and Wallenberg Foundation, Sweden; SERI, SNSF and Cantons of Bern and Geneva, Switzerland; NSTC, Taipei; TENMAK, T\"urkiye; STFC/UKRI, United Kingdom; DOE and NSF, United States of America.

Individual groups and members have received support from BCKDF, CANARIE, CRC and DRAC, Canada; CERN-CZ, FORTE and PRIMUS, Czech Republic; COST, ERC, ERDF, Horizon 2020, ICSC-NextGenerationEU and Marie Sk{\l}odowska-Curie Actions, European Union; Investissements d'Avenir Labex, Investissements d'Avenir Idex and ANR, France; DFG and AvH Foundation, Germany; Herakleitos, Thales and Aristeia programmes co-financed by EU-ESF and the Greek NSRF, Greece; BSF-NSF and MINERVA, Israel; NCN and NAWA, Poland; La Caixa Banking Foundation, CERCA Programme Generalitat de Catalunya and PROMETEO and GenT Programmes Generalitat Valenciana, Spain; G\"{o}ran Gustafssons Stiftelse, Sweden; The Royal Society and Leverhulme Trust, United Kingdom.

In addition, individual members wish to acknowledge support from Armenia: Yerevan Physics Institute (FAPERJ); CERN: European Organization for Nuclear Research (CERN PJAS); Chile: Agencia Nacional de Investigaci\'on y Desarrollo (FONDECYT 1230812, FONDECYT 1230987, FONDECYT 1240864); China: Chinese Ministry of Science and Technology (MOST-2023YFA1605700), National Natural Science Foundation of China (NSFC - 12175119, NSFC 12275265, NSFC-12075060); Czech Republic: Czech Science Foundation (GACR - 24-11373S), Ministry of Education Youth and Sports (FORTE CZ.02.01.01/00/22\_008/0004632), PRIMUS Research Programme (PRIMUS/21/SCI/017); EU: H2020 European Research Council (ERC - 101002463); European Union: European Research Council (ERC - 948254, ERC 101089007), Horizon 2020 Framework Programme (MUCCA - CHIST-ERA-19-XAI-00), European Union, Future Artificial Intelligence Research (FAIR-NextGenerationEU PE00000013), Italian Center for High Performance Computing, Big Data and Quantum Computing (ICSC, NextGenerationEU); France: Agence Nationale de la Recherche (ANR-20-CE31-0013, ANR-21-CE31-0013, ANR-21-CE31-0022, ANR-22-EDIR-0002), Investissements d'Avenir Labex (ANR-11-LABX-0012); Germany: Baden-Württemberg Stiftung (BW Stiftung-Postdoc Eliteprogramme), Deutsche Forschungsgemeinschaft (DFG - 469666862, DFG - CR 312/5-2); Italy: Istituto Nazionale di Fisica Nucleare (ICSC, NextGenerationEU), Ministero dell'Università e della Ricerca (PRIN - 20223N7F8K - PNRR M4.C2.1.1); Japan: Japan Society for the Promotion of Science (JSPS KAKENHI JP22H01227, JSPS KAKENHI JP22H04944, JSPS KAKENHI JP22KK0227, JSPS KAKENHI JP23KK0245); Netherlands: Netherlands Organisation for Scientific Research (NWO Veni 2020 - VI.Veni.202.179); Norway: Research Council of Norway (RCN-314472); Poland: Ministry of Science and Higher Education (IDUB AGH, POB8, D4 no 9722), Polish National Agency for Academic Exchange (PPN/PPO/2020/1/00002/U/00001), Polish National Science Centre (NCN 2021/42/E/ST2/00350, NCN OPUS nr 2022/47/B/ST2/03059, NCN UMO-2019/34/E/ST2/00393, UMO-2020/37/B/ST2/01043, UMO-2021/40/C/ST2/00187, UMO-2022/47/O/ST2/00148, UMO-2023/49/B/ST2/04085, UMO-2023/51/B/ST2/00920); Slovenia: Slovenian Research Agency (ARIS grant J1-3010); Spain: Generalitat Valenciana (Artemisa, FEDER, IDIFEDER/2018/048), Ministry of Science and Innovation (MCIN \& NextGenEU PCI2022-135018-2, MICIN \& FEDER PID2021-125273NB, RYC2019-028510-I, RYC2020-030254-I, RYC2021-031273-I, RYC2022-038164-I), PROMETEO and GenT Programmes Generalitat Valenciana (CIDEGENT/2019/027); Sweden: Swedish Research Council (Swedish Research Council 2023-04654, VR 2018-00482, VR 2022-03845, VR 2022-04683, VR 2023-03403, VR grant 2021-03651), Knut and Alice Wallenberg Foundation (KAW 2018.0157, KAW 2018.0458, KAW 2019.0447, KAW 2022.0358); Switzerland: Swiss National Science Foundation (SNSF - PCEFP2\_194658); United Kingdom: Leverhulme Trust (Leverhulme Trust RPG-2020-004), Royal Society (NIF-R1-231091); United States of America: U.S. Department of Energy (ECA DE-AC02-76SF00515), Neubauer Family Foundation.

%
%


%

%
%
%
\printbibliography

\clearpage
 
\begin{flushleft}
\hypersetup{urlcolor=black}
{\Large The ATLAS Collaboration}

\bigskip

\AtlasOrcid[0000-0002-6665-4934]{G.~Aad}$^\textrm{\scriptsize 104}$,
\AtlasOrcid[0000-0001-7616-1554]{E.~Aakvaag}$^\textrm{\scriptsize 17}$,
\AtlasOrcid[0000-0002-5888-2734]{B.~Abbott}$^\textrm{\scriptsize 123}$,
\AtlasOrcid[0000-0002-0287-5869]{S.~Abdelhameed}$^\textrm{\scriptsize 119a}$,
\AtlasOrcid[0000-0002-1002-1652]{K.~Abeling}$^\textrm{\scriptsize 56}$,
\AtlasOrcid[0000-0001-5763-2760]{N.J.~Abicht}$^\textrm{\scriptsize 50}$,
\AtlasOrcid[0000-0002-8496-9294]{S.H.~Abidi}$^\textrm{\scriptsize 30}$,
\AtlasOrcid[0009-0003-6578-220X]{M.~Aboelela}$^\textrm{\scriptsize 45}$,
\AtlasOrcid[0000-0002-9987-2292]{A.~Aboulhorma}$^\textrm{\scriptsize 36e}$,
\AtlasOrcid[0000-0001-5329-6640]{H.~Abramowicz}$^\textrm{\scriptsize 154}$
% ...
\bigskip
\\

% ...
$^{17}$Department for Physics and Technology, University of Bergen, Bergen; Norway.\\
% ...
$^{30}$Physics Department, Brookhaven National Laboratory, Upton NY; United States of America.\\
% ...
$^{36}$$^{(a)}$Facult\'e des Sciences Ain Chock, Universit\'e Hassan II de Casablanca;$^{(b)}$Facult\'{e} des Sciences, Universit\'{e} Ibn-Tofail, K\'{e}nitra;$^{(c)}$Facult\'e des Sciences Semlalia, Universit\'e Cadi Ayyad, LPHEA-Marrakech;$^{(d)}$LPMR, Facult\'e des Sciences, Universit\'e Mohamed Premier, Oujda;$^{(e)}$Facult\'e des sciences, Universit\'e Mohammed V, Rabat;$^{(f)}$Institute of Applied Physics, Mohammed VI Polytechnic University, Ben Guerir; Morocco.\\
% ...
$^{45}$Physics Department, Southern Methodist University, Dallas TX; United States of America.\\
% ...
$^{50}$Fakult\"{a}t Physik , Technische Universit{\"a}t Dortmund, Dortmund; Germany.\\
% ...
$^{56}$II. Physikalisches Institut, Georg-August-Universit\"{a}t G\"ottingen, G\"ottingen; Germany.\\
% ...
$^{104}$CPPM, Aix-Marseille Universit\'e, CNRS/IN2P3, Marseille; France.\\
% ...
$^{119}$$^{(a)}$New York University Abu Dhabi, Abu Dhabi;$^{(b)}$United Arab Emirates University, Al Ain; United Arab Emirates.\\
% ...
$^{123}$Homer L. Dodge Department of Physics and Astronomy, University of Oklahoma, Norman OK; United States of America.\\
% ...
$^{154}$Raymond and Beverly Sackler School of Physics and Astronomy, Tel Aviv University, Tel Aviv; Israel.\\
% ...
$^{a}$ Also Affiliated with an institute covered by a cooperation agreement with CERN.\\
$^{b}$ Also at An-Najah National University, Nablus; Palestine.\\
$^{c}$ Also at Borough of Manhattan Community College, City University of New York, New York NY; United States of America.\\
$^{d}$ Also at Center for Interdisciplinary Research and Innovation (CIRI-AUTH), Thessaloniki; Greece.\\
$^{e}$ Also at CERN, Geneva; Switzerland.\\
$^{f}$ Also at CMD-AC UNEC Research Center, Azerbaijan State University of Economics (UNEC); Azerbaijan.\\
% ...

\end{flushleft}


\end{document}